\documentclass{article}
\usepackage{multirow}
\usepackage{amsmath}
\usepackage{graphicx} % Required for inserting images
\usepackage{float}
\usepackage[a4paper, margin=1in]{geometry}
\usepackage{hyperref}
\usepackage{pdfpages}
\usepackage{cite}
\usepackage[utf8]{inputenc}
\usepackage{hyphenat}
\usepackage{authblk}

\title{Edge-state competition in a 2D topological insulator-semiconductor heterostructure}
\author[1,2]{Wei Li}
\author[3]{Pier Philipsen}
\author[1]{Thomas Brumme}
\author[1,2,4]{Thomas Heine}

\affil[1]{TU Dresden, Theoretical Chemistry, Bergstr. 66c, 01062 Dresden, Germany}
\affil[2]{CASUS -- Center for Advanced Systems Understanding, Helmholtz-Zentrum Dresden-Rossendorf e.V. (HZDR), Untermarkt 20, D-02826~Görlitz, Germany}
\affil[3]{Software for Chemistry \& Materials BV, De Boelelaan 1109, 1081HV Amsterdam, The Netherlands}
\affil[4]{Yonsei University and ibs-cnm, Seodaemun-gu, Seoul 120-749, Republic of Korea}

\date{}

\begin{document}

\maketitle
\noindent Email: thomas.heine@tu-dresden.de

\section{Abstract}
Quantum spin Hall edge transport in two-dimensional transition-metal dichalcogenides depends on whether their one-dimensional edge channels are preserved under realistic substrates and device boundaries. 
Here we implement spin-orbit coupling in DFTB and GFN-xTB within the Amsterdam Modeling Suite, and apply it to 1T$'$/2H WSe$_2$ heterostructures. 
Edge-projected spectra reveal robust edge states in 1T$'$ ribbons; and these states remain robust against a laterally infinite 2H substrate, which only shifts the Dirac point via long-wavelength corrugation without introducing additional in-gap states. By contrast, terminated 2H edges generate trivial dispersion branches in the same energy window that hybridize only weakly with the topological edge modes. In the bulk, Fermi-level states are 1T$'$-derived; at the small twist angle, lattice-relaxation-induced strain drives miniband reconstruction, whereas at the large twist angle, the layers become electronically decoupled. These findings suggest the conditions -- controlled twist angle and avoidance of terminated 2H edges -- for achieving quantized conductance and unambiguous spectroscopic detection of helical edges.

\section{Introduction}
Quantum spin Hall (QSH) insulators\cite{Kane2005,Kane2005-2,Bernevig2006,Qi2011} have an insulating bulk but conducting edge states that are topologically protected from backscattering by time-reversal symmetry, holding promise for quantum electronic devices with low dissipation.
Monolayer 1T$'$-WSe$_2$ has emerged as a particularly notable QSH. Theory predicted band inversion and a spin-orbit coupling (SOC) induced gap opening in monolayer 1T$'$-MX$_2$ (M = Mo or W; X = S, Se, Te), and experiments have confirmed its synthesis and revealed a topological band gap on the order of 0.1-0.2 eV\cite{Chen2018}.

The semiconducting 2H polytype of WSe$_2$ can serve as a substrate, thereby interfacing a topological insulator with a trivial semiconductor. Such a platform also enables coupling between the spin-valley physics of the 2H layer and the topological states of the 1T$'$ layer.
However, building such heterostructures introduces two central complications.
First, in the bilayer systems, the control of the interlayer rotation introduces a new degree of freedom -- twist angle --for tuning electronic properties\cite{Bistritzer2011,Gong2014,Naik2018-pl,Liu2021-gb,Tran2019-ct,Angeli2021-zf}. 
Second, it is well-known that edges can host metallic bands -- as reported, for example, for MoS$_2$\cite{Bollinger2001}. When present inside a device, these trivial edge bands may obscure or hybridize with the desired topological channels.

It therefore remains unclear whether the 1T$'$ layer retains its topological character once it is strained by the 2H substrate and whether interlayer hybridization or proximate trivial edges can destroy or mask the QSH edge modes.
Furthermore, in realistic, finite-sized devices, it remains unexplored how the 1T$'$ layer's signature topological edge states can be distinguished from the conventional, trivial edge states that may form at the boundaries of the 2H semiconductor.

To address these questions, we focus on edgeless 1T$'$/2H WSe$_2$ heterostructures across twist angles so that only bulk-to-bulk interactions and structural relaxation influence the band structures. 
Then we turn to nanoribbon heterostructures -- a commonly used approach for resolving topological edge modes because a ribbon geometry exposes the one-dimensional edge dispersions explicitly in the band structure\cite{Ma2016,Ma20162,Ma2018}.
A finite 1T$'$ strip is contacted either by an extended 2H sheet or by a slightly wider 2H ribbon. These setups allow us to disentangle the roles of interlayer hybridization, structural relaxation, and edge proximity.

Moiré superlattices and nanoribbons are computationally expensive because they contain hundreds to thousands of atoms.
To address this challenge in large-scale simulations, semi-empirical methods such as the Density-Functional based Tight Binding (DFTB)\cite{Porezag1995,Seifert1996,Elstner1998,Porezag1995,Yang2007,Gaus2011} and the Geometry, Frequency, and Noncovalent interaction eXtended Tight-Binding (GFN-xTB) family\cite{Grimme2017,Bannwarth2019} have been developed.
K\"ohler \textit{et~al.} formulated SOC within the DFTB framework~\cite{Kohler2007}. Recently, Jha \textit{et al.} proposed a set of spin-orbit parameters covering nearly the entire periodic table ($1\le Z\le 118$)~\cite{Jha2022} and validated it within DFTB~\cite{Jha2022} and GFN1-xTB~\cite{Jha2023} calculations. These developments demonstrate that semi-empirical methods can be systematically enhanced to treat relativistic effects, broadening their applicability to heavy-element systems and spin-dependent phenomena.
In this work, we introduce our implementation of SOC for both DFTB and GFN-xTB within the Amsterdam Modeling Suite (AMS). While the open-source DFTB+ program\cite{Hourahine2020} has SOC functionality, AMS offers unique advantages: it provides DFTB1, DFTB2, DFTB3, and GFN1-xTB engines that share the same infrastructure with its DFT modules. This tight integration enables hybrid simulations that seamlessly mix DFT and DFTB levels of theory within a single calculation. We have implemented SOC terms in the AMS DFTB module for both DFTB and the GFN1-xTB Hamiltonian~\cite{Kohler2007,Jha2022}. Our implementation builds on the theoretical approach of adding a spin-orbit two-component Hamiltonian to the DFTB formalism~\cite{Kohler2007}, using the published atomic SOC parameters~\cite{Jha2022}, and is validated through benchmark calculations. The same strategy can be applied to other GFN-xTB variants or future semi-empirical models that require relativistic corrections.

In the following, we detail the implementation of SOC in the DFTB and GFN1-xTB frameworks and present benchmarks in Supporting Information validating its accuracy. We then apply the method to 1T$'$/2H-WSe$_2$ heterostructures to present the miniband reconstruction at the small twist angle and the interlayer decoupling at the large twist angle; next we analyze edge spectra and hybridization in finite nanoribbons.
\section{Method}
\subsection{The DFTB formulation}
The DFTB method, primarily pioneered by Seifert, Elstner, Frauenheim and coworkers\cite{Porezag1995,Seifert1996,Elstner1998,Yang2007,Gaus2011}, can be derived from the Kohn-Sham formulation of DFT\cite{KohnSham1965}. In DFTB, the total energy of the system is expanded as a Taylor series with respect to a reference density $\rho_0$ (the superposition of neutral atomic densities):
\begin{equation}
\begin{split}
    E_{\text{tot}}[\rho_0 + \delta\rho] &= E^0[\rho_0] + E^1[\rho_0, \delta\rho] + E^2[\rho_0, \delta\rho^2] + E^3[\rho_0, \delta\rho^3] +\mathcal{O}(\delta\rho^4) \\
    &= \sum_{I < J} V^{\text{rep}}_{IJ}(R_{IJ}) + \sum_i^{\text{occ}} f_i \bigl\langle \psi_i \bigl| \hat{H}[\rho_{0}] \bigr| \psi_i \bigr\rangle \\
    & \qquad + \frac{1}{2} \int \int \left[
        \frac{1}{|\boldsymbol{r}-\boldsymbol{r}'|} + \left. \frac{\delta^2 E_{\text{xc}}}{\delta \rho(\boldsymbol{r}) \delta \rho(\boldsymbol{r}')} \right|_{\rho_0} \right] 
        \delta \rho(\boldsymbol{r}) \delta \rho(\boldsymbol{r}') \, d\boldsymbol{r} \, d\boldsymbol{r}' \\
    & \qquad + \frac{1}{6} \int \int \int 
        \left. \frac{\delta^3 E_{\text{xc}}}{\delta \rho(\boldsymbol{r}) \delta \rho(\boldsymbol{r}' ) \rho(\boldsymbol{r}'')} \right|_{\rho_0}
       \delta \rho(\boldsymbol{r}) \delta \rho(\boldsymbol{r}') \delta \rho(\boldsymbol{r}'') \, d\boldsymbol{r} \, d\boldsymbol{r}'\,d\boldsymbol{r}'' +\mathcal{O}(\delta\rho^4).
\end{split}
\label{eq:KS}
\end{equation}
Here $E_{\text{xc}}$ is the exchange correlation (XC) energy.

Following the Linear Combination of Atomic Orbitals (LCAO) ansatz, each molecular orbital $\psi_i$ is expanded in a minimal basis set:
\begin{equation}
    \psi_i = \sum_{\mu} c_{\mu i} \phi_{\mu}.
\end{equation}
We then introduce the charge fluctuations by decomposing the density fluctuations $\delta\rho$ into atomic contributions:
\begin{equation}
    \delta\rho(\boldsymbol{r}) = \sum_I \Delta q_I \delta\rho_I(\boldsymbol{r}),
\end{equation}
where the atomic charge fluctuations $\Delta q_I$ are evaluated within Mulliken population analysis\cite{Mulliken1955}.
Thus, the total energy of DFTB1\cite{Porezag1995} is given by:
\begin{equation}
\begin{split}
    E_{\text{DFTB1}} &= E^{\text{rep}} + E^{H^0} \\
    &= \sum_{I < J} V^{\text{rep}}_{IJ}(R_{IJ}) + \sum_{i}^{\text{occ}}f_{i}
      \sum_{I,J}\sum_{\mu\in I}\sum_{\nu\in J}c_{\mu i}c_{\nu i}\,H^{0}_{\mu\nu}.
\end{split}
\end{equation}
Here  $V^{\text{rep}}_{IJ}$ is a pairwise repulsive potential between atoms $I$ and $J$. The second term is the band-structure energy, where $H^{0}_{\mu\nu} = \langle\phi_{\mu}|\hat{H}_{\mu\nu}^{0}|\phi_{\nu}\rangle$ is given within the LCAO scheme as an approximation to the Kohn-Sham Hamiltonian $\hat{H}[\rho_0]$ from Eq.~\ref{eq:KS}.

The DFTB2\cite{Seifert1996,Elstner1998} method accounts for charge redistribution within the system:
\begin{equation}
\begin{split}
    E^{\text{DFTB2}} &= E_{\text{DFTB1}} + E^\gamma \\
    &= E^{\text{rep}} + \sum_{i}^{\text{occ}}f_{i}
      \sum_{I,J}\sum_{\mu\in I}\sum_{\nu\in J}c_{\mu i}c_{\nu i}\,H^{0}_{\mu\nu} + 
    \frac{1}{2} \sum_{I J} \Delta q_I \Delta q_J \gamma_{IJ},
\end{split}
\end{equation}Writing the wave function of the systems as two-component spinors instead of scalar wavefunctions, the total electronic and magnetization density can be given as the linear combination of the Pauli matrices
where $\gamma_{IJ}$ function is given by the integral over a product of two normalized Slater-type spherical charge densities, and depends on the Hubbard parameters and the covalent radius.

The DFTB3\cite{Yang2007,Gaus2011} method extends DFTB2 by including the third-order contribution:
\begin{equation}
\begin{split}
    E^{\text{DFTB3}} &= E^{\text{DFTB2}} + E^\Gamma \\
    &= E^{\text{rep}} + \sum_{i}^{\text{occ}}f_{i}
      \sum_{I,J}\sum_{\mu\in I}\sum_{\nu\in J}c_{\mu i}c_{\nu i}\,H^{0}_{\mu\nu} + 
    \frac{1}{2} \sum_{I J} \Delta q_I \Delta q_J \gamma_{IJ} + 
    \frac{1}{3} \sum_{I J} \Delta q_I^2 \Delta q_J \Gamma_{IJ},
\end{split}
\end{equation}
where $\Gamma_{IJ}$ is a derivative of the $\gamma_{IJ}$ function with respect to the charge fluctuations.

Consequently, the Hamiltonian matrix elements $H_{\mu\nu} = \langle \phi_{\mu} | \hat{H}_{\mu\nu} | \phi_{\nu} \rangle$ are given by:
\begin{align}
    H_{\mu\nu} &= H_{\mu\nu}^{0} 
               + H_{\mu\nu}^{2}[\gamma,\Delta q] 
               + H_{\mu\nu}^{3}[\Gamma,\Delta q],
               \qquad \mu\!\in\! I, \; \nu\!\in\! J, \\
    H_{\mu\nu}^{2} &= \frac{S_{\mu\nu}}{2} 
        \sum_{K} \bigl(\gamma_{JK} + \gamma_{IK}\bigr) \, \Delta q_{K}, \\
    H_{\mu\nu}^{3} &= S_{\mu\nu} \sum_{K} \Delta q_K 
        \left[ \frac{1}{3} \bigl( \Delta q_I \Gamma_{IK} + \Delta q_J \Gamma_{JK} \bigr) 
             + \frac{\Delta q_K}{6} \bigl( \Gamma_{IK} + \Gamma_{JK} \bigr) \right].
\end{align}

\subsection{The GFN-xTB formulation}

The GFN-xTB family of methods\cite{Grimme2017} shares a similar theoretical foundation with DFTB, extending the total energy expression with additional empirical terms. The total energy in GFN1-xTB is decomposed as:
\begin{equation}
E_{\text{tot}} = E_{\text{el}} + E_{\text{rep}} + E_{\text{disp}} + E_{\text{XB}}.
\end{equation}
Here, $E_{\text{rep}}$ is a pairwise repulsion, while $E_{\text{disp}}$ and $E_{\text{XB}}$ are corrections for dispersion and halogen bonding, respectively.

The electronic energy $E_{\text{el}}$ is formulated similarly to DFTB3, including band-structure, second-order, and third-order contributions, plus an electronic entropy term for finite-temperature occupations:
\begin{equation}
\begin{split}
E_{\text{el}} &= E^{H^0} + E^2 + E^3 + E_{\text{el}}^{\text{ent}} \\
&= \sum_{i}^{\text{occ}}f_{i} \sum_{I,J}\sum_{\mu\in I}\sum_{\nu\in J}c_{\mu i}c_{\nu i}\,H^{0}_{\mu\nu}
+ \frac{1}{2} \sum_{I,J} \sum_{l \in I} \sum_{l' \in J} q_l q_{l'} \gamma_{IJ,ll'} +  \frac{1}{3} \sum_I \Gamma_I q_I^3 + E_{\text{el}}^{\text{ent}}.
\end{split}
\end{equation}
The first term, the band-structure energy $E^{H^0}$, is identical in form to the DFTB expression.
The second-order term $E^2$ describes the isotropic electrostatic and XC energy which originates from the second-order term in the energy expansion, with $q_l$ being partial Mulliken shell charges and $\gamma$ being a short-ranged damped Coulomb interaction. 
The third-order term $E^3$, different from DFTB3, is an on-site electrostatic/XC correction, where $\Gamma_I$ is the charge derivative of the Hubbard parameter $U$ and $q_I$ is the atomic Mulliken charge.
The last term $E_{\text{el}}^{\text{ent}}$ is the electronic entropy, required for finite-temperature Fermi occupations. $I$ and $J$ are two distinct atoms of the system, $l$ and $l'$ are the orbital angular momentum of the atomic shells of atoms $I$ and $J$, respectively.

Thus the matrix elements of the GFN-xTB Hamiltonian $H_{\mu\nu} = \langle \phi_{\mu} | \hat{H}_{\mu\nu} | \phi_{\nu} \rangle$ can be expressed as:
\begin{equation}
\begin{split}
    H_{\mu\nu} &= H_{\mu\nu}^{0} + H_{\mu\nu}^{2} + H_{\mu\nu}^{3},  \qquad (\mu\!\in\! l(I),\; \nu\!\in\! l'(J)), \\
    H_{\mu\nu}^{2} &= \frac{1}{2}\,S_{\mu\nu}
        \sum_{K}\sum_{l''\in K}\!
          \bigl(\gamma_{IK,ll''} + \gamma_{JK,l'l''}\bigr)\,
          q_{l''}^{K}, \\
    H_{\mu\nu}^{3} &= \frac{1}{2}\,S_{\mu\nu}
        \bigl(q_{I}^{2}\Gamma_{I} + q_{J}^{2}\Gamma_{J}\bigr).
\end{split}
\end{equation}

\subsection{Spin-orbit coupling}
SOC is a relativistic term originating intrinsically from the solution of the Dirac equation. In the non-relativistic limit\cite{Foldy1950}, SOC appears as an additional term in the Schr\"odinger equation given by:
\begin{equation}
    \hat{H}_{\text{DIRAC-SOC}} = - \frac{e \hbar}{4m^2 c^2} \hat{\boldsymbol{\sigma}} \cdot (\boldsymbol{E} \times \hat{\boldsymbol{p}}),
\end{equation}
where $\hat{\boldsymbol{\sigma}}$ are 2 $\times$ 2 Pauli matrices, and $\hat{\boldsymbol{S}} = \frac{\hbar}{2}\hat{\boldsymbol{\sigma}}$ is the spin operator. 
The electric field of the nucleus, $\boldsymbol{E}$, is approximated by a spherically symmetric atomic potential $\boldsymbol{E} = -\boldsymbol{\nabla} V(r)/e = -\frac{1}{e} \frac{dV}{dr}\boldsymbol{e}_{r}$. Consequently, the SOC operator can be rewritten in the well-known form that couples the electron's spin and orbital angular momentum:
\begin{equation}
\hat{H}_\text{SO} = \zeta (r) \hat{\boldsymbol{L}} \cdot \hat{\boldsymbol{S}},
\end{equation}
where $\hat{\boldsymbol{L}}$ is the orbital angular momentum operator, $\hat{\boldsymbol{S}}$ is the spin angular momentum operator, and $\zeta(r)=\frac{1}{2m^{2}c^{2}r}\frac{dV}{dr}$ describes the SOC strength.

In the DFTB and GFN-xTB frameworks, SOC is incorporated as an on-site contribution to the Hamiltonian\cite{Kohler2007}, whereby only the $\langle \phi_{\mu} | \hat{H}_{\text{SO}} | \phi_{\nu} \rangle$ matrix elements with $\mu$ and $\nu$ belonging to the same atom are considered, while the rest of the interatomic hopping terms are discarded. The  SOC operator is approximated with a set of constant, element- and shell-dependent SOC parameters $\xi_I^l$:
\begin{equation}
\hat{H}_{\text{SO},I}^l = \frac{\xi^l_I}{2} \begin{pmatrix}
\hat{L}_z & \hat{L}_- \\
\hat{L}_+ & -\hat{L}_z
\end{pmatrix}.
\end{equation}
Here $\hat{L}_z$, $\hat{L}_\pm$ are the $z$-component and ladder operators of the orbital angular momentum, respectively, whose actions on the spherical harmonics are:
$
\hat{L}_z |l,m \rangle = \hbar m |l,m \rangle$, \\ 
$\hat{L}_\pm |l,m \rangle = \hbar \sqrt{l(l + 1) - m(m \pm 1)} |l,m \pm 1 \rangle
$.
Molecular orbitals are therefore expressed as two-component spinors:
\begin{equation}
\begin{pmatrix}
\psi_i^\alpha \\
\psi_i^\beta
\end{pmatrix} = \sum_\mu \begin{pmatrix}
c_{\mu i}^\alpha \\
c_{\mu i}^\beta
\end{pmatrix} \varphi_\mu,
\end{equation}
where $\varphi_\mu$ are the atomic-like basis functions, and $c_{\mu i}^{\sigma}$ the spin-dependent coefficients ($\sigma\!\in\!{\alpha,\beta}$).  For periodic systems an additional crystal momentum index $\boldsymbol{k}$ is appended (see below).

By adding the on-site SOC contribution to the scalar-relativistic part $H_{\mu\nu}^{\text{SR}}=H_{\mu\nu}^0+H_{\mu\nu}^2+H_{\mu\nu}^3$ inherited from either DFTB or GFN-xTB, the total non-collinear Hamiltonian matrix is:
\begin{align}
H_{\mu\nu} &= \begin{pmatrix}
H^{\alpha\alpha}_{\mu\nu} & H^{\alpha\beta}_{\mu\nu} \\
H^{\beta\alpha}_{\mu\nu} & H^{\beta\beta}_{\mu\nu}
\end{pmatrix} \\
&= H_{\mu\nu}^{\text{SR}} \otimes \begin{pmatrix}
1 & 0 \\
0 & 1
\end{pmatrix}
+ \frac{S_{\mu\nu}}{2}\left[\xi_I^l \begin{pmatrix}
\hat{L}_z & \hat{L}_- \\
\hat{L}_+ & -\hat{L}_z
\end{pmatrix}_l + \xi_J^{l'} \begin{pmatrix}
\hat{L}_z & \hat{L}_- \\
\hat{L}_+ & -\hat{L}_z
\end{pmatrix}_{l'}\right],
\label{eq:fullHam}
\end{align}
where  $\sigma,\sigma'\in\{\alpha,\beta\}$, and $\mu\!\in\! l(I),\; \nu\!\in\! l'(J)$. $S_{\mu \nu} \equiv \langle \varphi_{\mu} | \varphi_{\nu} \rangle$ is the overlap matrix.
The spinor coefficients $c_{\mu i}$ and corresponding orbital energies $\epsilon_i$ are then calculated by solving the secular equation:
\begin{equation}
\begin{split}
    \sum_{\mu} \left[
    \begin{pmatrix}
    H^{\alpha\alpha}_{\mu\nu}
    &H^{\alpha\beta}_{\mu\nu} \\
    H^{\beta\alpha}_{\mu\nu} 
    &H^{\beta\beta}_{\mu\nu}
    \end{pmatrix}  - \epsilon_{i} S_{\mu\nu} \begin{pmatrix} 1 & 0 \\ 0 & 1 \end{pmatrix} \right]
    \begin{pmatrix}
    c^{\alpha}_{\mu i} \\
    c^{\beta}_{\mu i}
    \end{pmatrix}
    = 0.
\end{split}
\end{equation}

\subsubsection{Density matrix and observables}
Writing the molecular orbitals as two-component spinors instead of scalar wavefunctions, the total electronic and magnetization density can be given as the linear combination of the Pauli matrices:
\begin{equation}
\begin{split}
\boldsymbol{\rho}(\boldsymbol{r}) &= \sum_i^{\text{occ}} f_i \begin{pmatrix}
\psi_i^{\alpha *} \\ \psi_i^{\beta *}
\end{pmatrix} \begin{pmatrix}
\psi_i^{\alpha} \\ \psi_i^{\beta}
\end{pmatrix} = \begin{pmatrix}
\rho^{\alpha\alpha} & \rho^{\alpha\beta} \\
\rho^{\beta\alpha} & \rho^{\beta\beta}
\end{pmatrix} \\
&= n(\boldsymbol{r}) \begin{pmatrix}
    1 & 0 \\
    0 & 1
\end{pmatrix} + 
m^x(\boldsymbol{r}) \begin{pmatrix}
    0 & 1 \\
    1 & 0
\end{pmatrix} + 
m^y(\boldsymbol{r}) \begin{pmatrix}
    0 & -i \\
    i & 0
\end{pmatrix} + 
m^z(\boldsymbol{r}) \begin{pmatrix}
    1 & 0 \\
    0 & -1
\end{pmatrix}, 
\end{split}
\end{equation}

Consequently, we obtain:
\begin{align}
n(\boldsymbol{r}) &= \frac{1}{2}  \left(\rho^{\alpha\alpha} + \rho^{\beta\beta}\right), \\
\boldsymbol{m}(\boldsymbol{r}) &= \bigl(\frac{1}{2}
(\rho^{\alpha\beta} + \rho^{\beta\alpha}),
-\frac{i}{2}(\rho^{\alpha\beta} - \rho^{\beta\alpha}),
\frac{1}{2}(\rho^{\alpha\alpha} - \rho^{\beta\beta})
\bigr).
\end{align}
The density components in the spinor space $\rho_{\mu\nu}^{\sigma\sigma'}$ are related to the density matrix through $\rho_{\mu\nu}^{\sigma\sigma'}=P_{\mu\nu}^{\sigma\sigma'} S_{\mu\nu}$ where  $P_{\mu\nu}^{\sigma\sigma'}=\sum_i f_i  c_{\mu i}^{\sigma} c_{\nu i}^{\sigma',*}$ represents the spinor-resolved density matrix.
Extending the standard Mulliken population analysis to the spinor formalism gives the following expression for the Mulliken charge on atom I in the presence of SOC:
\begin{equation}
q_I = \frac{1}{2}\left(q_I^{\alpha\alpha} + q_I^{\beta\beta}\right) 
= \frac{1}{2}\sum_{\mu \in I} \sum_\nu \left(\rho_{\mu\nu}^{\alpha\alpha} + \rho_{\mu\nu}^{\beta\beta} \right)S_{\mu\nu}.
\end{equation}

\subsubsection{Energy and forces}
The spin-orbit contribution to the total energy can be calculated as:
\begin{equation}
E_\text{SO} = \operatorname{Tr} (P H_\text{SO}) = \sum_{\mu\nu} \left[H_\text{SO}^{\alpha\alpha} P^{\alpha\alpha}_{\nu\mu} + H_\text{SO}^{\beta\beta} P^{\beta\beta}_{\nu\mu} + 2 \operatorname{Re}\left(H_\text{SO}^{\alpha\beta} P^{\alpha\beta}_{\nu\mu}\right)\right].
\end{equation}

The spin-orbit contribution to the forces on atoms can be written as:
\begin{equation}
\boldsymbol{F}_\text{SO,I} = -\sum_{\sigma\sigma'} \frac{\partial E_\text{SO}^{\sigma\sigma'}}{\partial \boldsymbol{R}_I} = \sum_{\mu\nu,\boldsymbol{R}} \left[
\sum_{\sigma\sigma'} P^{\sigma\sigma'}_{\mu\nu}(\boldsymbol{R}) \frac{\partial H_\text{SO}^{\sigma\sigma'}(\boldsymbol{R})}{\partial \boldsymbol{R}_I}
- \sum_{\sigma\sigma'} P_{\mu\nu}^{\epsilon,\sigma\sigma'}(\boldsymbol{R}) \frac{\partial S_{\mu\nu}(\boldsymbol{R})}{\partial \boldsymbol{R}_I}
\right],
\end{equation}
where $P_{\text{SO}}^{\epsilon,\sigma\sigma'}$ denotes the energy-weighted density matrix,
$P_{\mu\nu}^{\epsilon,\sigma\sigma'}=\sum_i f_i \epsilon_i c_{\mu i}^{\sigma} c_{\nu i}^{\sigma',*}$.
Because on-site SOC integrals depend only on the local atomic potential, $\partial H^{\sigma\sigma'}_{\text{SO}}/\partial\boldsymbol R_I = 0$.
The overlap-related term vanishes as well since $\rho_{\text{SO}}^{\epsilon}$ is antisymmetric in $(\mu,\nu)$ whereas $S_{\mu\nu}$ is symmetric, making overlap contribution zero. Nevertheless, SOC affects the forces indirectly through the self-consistent modification of the charges.

\subsubsection{Periodic systems}
The implementation of SOC in DFTB and GFN-xTB can be extended to periodic systems by using Bloch wavefunctions:
\begin{equation}
\varphi_{\mu}(\boldsymbol{k}, \boldsymbol{r}) = \frac{1}{\sqrt{N}} \sum_{\boldsymbol{T}} e^{i \boldsymbol{k} \cdot \boldsymbol{T}} \varphi_{\mu}(\boldsymbol{r} - \boldsymbol{T}),
\end{equation}
which means that a basis function $\varphi_{\mu}(\boldsymbol{k}, \boldsymbol{r})$ changes by a phase $e^{i \boldsymbol{k} \cdot \boldsymbol{T}}$ in translation $\boldsymbol{T}$. In practice, the Brillouin zone is sampled on a discrete grid of k-points, for which the secular equation is solved.

This allows for the calculation of band structures and corresponding spin texture for materials exhibiting strong SOC. For example, the $\boldsymbol{k}$-resolved orbital projection (commonly visualized as so-called fatbands) is given by:
\begin{equation}
n_{i\boldsymbol{k}}^\mu = \frac{1}{2} \sum_\nu \left(c_{\mu i\boldsymbol{k}}^{\alpha *} c_{\nu i\boldsymbol{k}}^{\alpha} S_{\mu\nu} + c_{\mu i\boldsymbol{k}}^{\beta *} c_{\nu i\boldsymbol{k}}^{\beta} S_{\mu\nu}\right),
\end{equation}
and the spin texture along z as:
\begin{equation}
m_{i\boldsymbol{k}}^{\mu,z} = \frac{1}{2} \sum_\nu \left(c_{\mu i\boldsymbol{k}}^{\alpha *} c_{\nu i\boldsymbol{k}}^{\alpha} S_{\mu\nu} - c_{\mu i\boldsymbol{k}}^{\beta *} c_{\nu i\boldsymbol{k}}^{\beta} S_{\mu\nu}\right).
\end{equation}

\subsection{Computational details of the benchmark calculations}

Geometry optimizations were performed with the GFN1-xTB method as implemented in AMS. SOC was enabled by supplying the parameter set of Ref.\cite{Jha2022}, which was also employed in all DFTB calculations. Self-consistent-charge (SCC) cycles were considered converged when the total energy changed by less than $1\times10^{-8}$ Hartree between iterations. The electronic structure was evaluated both with and without SOC. Uniform Monkhorst-Pack\cite{Monkhorst1976-tk} $k$-point meshes were generated by sampling the Brillouin zone with a reciprocal-space spacing of $0.01\ \text{\AA}^{-1}$.

Benchmark calculations reported in the Supporting Information follow the same but additionally include SCC-DFTB single-point calculation using the {QUASINANO2013}\cite{Wahiduzzaman2013-sv} Slater-Koster parameters, again augmented by the SOC parameter set of Ref.~\cite{Jha2022}. 
Calculations were performed with AMS and DFTB+\cite{Hourahine2020} with identical Slater-Koster and SOC parameter files, $k$-meshes, and convergence thresholds. This duplication allows direct cross-validation of the AMS and DFTB+ implementations. 
For structures taken directly from the literature, the published geometries were retained and re-evaluated at both the xTB and DFTB levels described above.

\section{Results and Discussion}

\subsection{1T$'$ monolayer}
\begin{figure}[htbp]
    \centering
    \includegraphics[width=\textwidth]{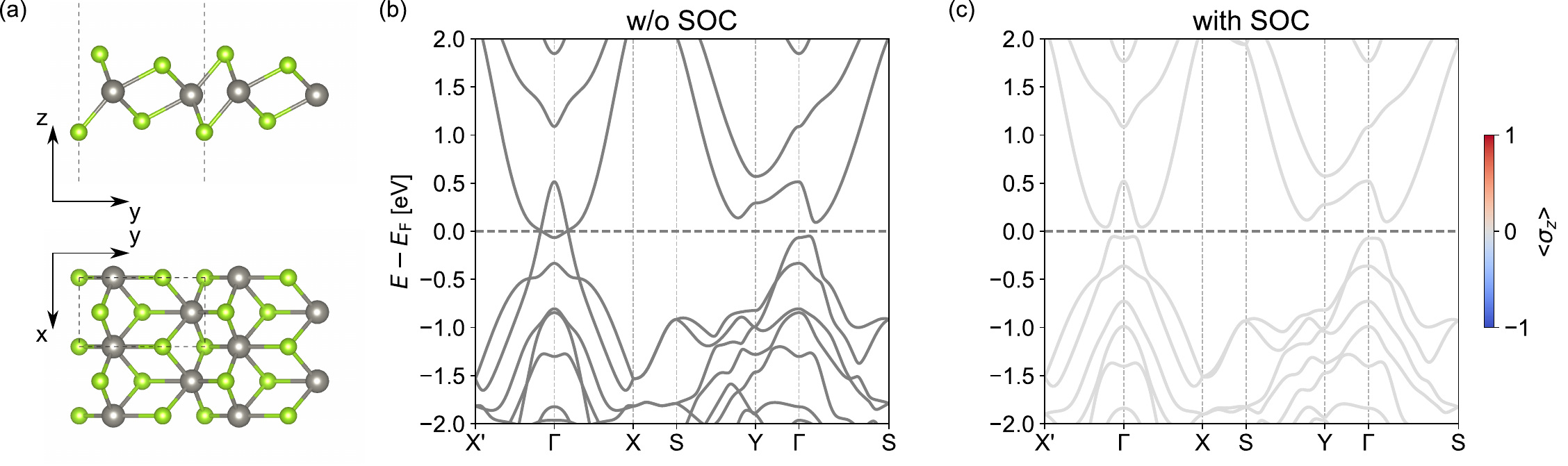}
    \caption{Electronic structure of a 1T$'$-WSe$_2$ monolayer. (a) Side (top) and top-down (bottom) views of the 1T$'$ crystal structure (W, grey; Se, green). (b) Band structure calculated without SOC, along the high-symmetry path X'-$\Gamma$-X-S-Y-$\Gamma$-S, showing a semi-metallic character with a Dirac-like crossing along the $\Gamma$-X direction. (c) Corresponding band structure calculated with SOC, showing that a band gap of 106~meV is opened near the $\Gamma$ point. The Fermi level is set to 0~eV.}
    \label{fig:mono}
\end{figure}

As a first step, we benchmarked our SOC-enabled GFN1-xTB method against established results for the 1T$'$-WSe$_2$ monolayer\cite{Qian14} to verify its accuracy in capturing fundamental electronic properties. 
In the absence of SOC, the material exhibits a semi-metallic band structure characterized by a band crossing along the $\Gamma$-X high-symmetry line (Figure~\ref{fig:mono} (b)), where the non-trivial band topology arises from the inverted ordering of W-5$d$ and Se-4$p$ states near the Fermi level\cite{Qian14,Chen2018}.
The inclusion of SOC opens a band gap of 106~meV at the former band-crossing point, transforming the material into a quantum spin Hall (QSH) insulator (Figure~\ref{fig:mono} (c)). This value is in good quantitative agreement with the 116~meV gap predicted by previous GW calculations\cite{Qian14}. This successful benchmark allows us to explore much larger and more complex moiré superlattices in the following.

\subsection{1T$'$/2H high-symmetry stacking}

The 1T$'$/2H WSe\(_2\) heterostructure combines two distinct crystalline phases. The 2H phase exhibits the thermodynamically stable trigonal prismatic coordination with $\bar{6}m2$ point group symmetry, while the 1T$'$ phase represents a distorted octahedral structure arising from a $2 \times 1$ structural distortion of the parent 1T phase. The distortion forms zigzag W-W chains along one in-plane direction, lowering the symmetry to $2/m$ and producing an orthorhombic unit cell.

To understand interlayer coupling, we begin by studying a commensurate heterostructure with high-symmetry stacking. Such a simple configuration allows for the study of intrinsic electronic hybridization without the structural complexities introduced by a moiré pattern.

\begin{figure}[htbp]
    \centering
    \includegraphics[width=\textwidth]{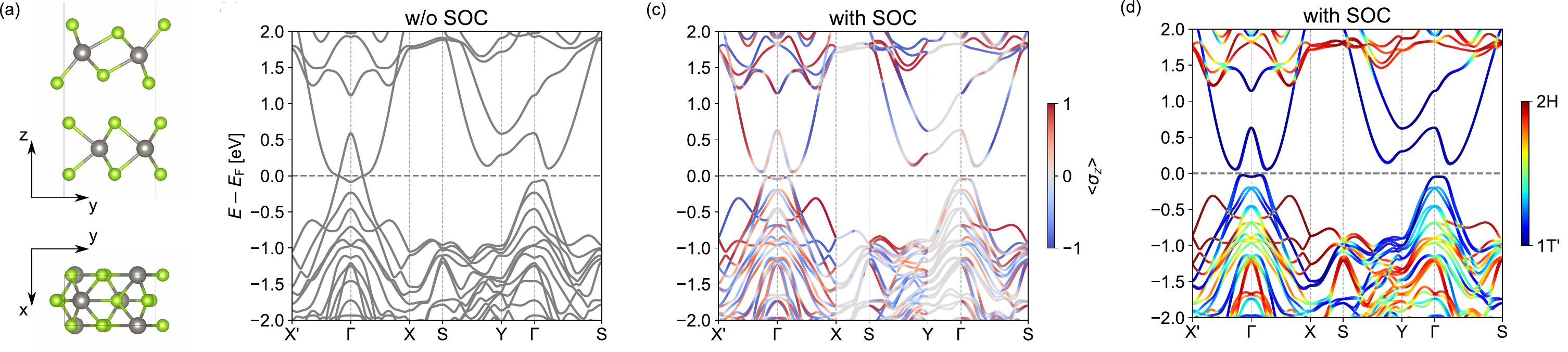}
    \caption{Atomic structure of a $Mh$ stacking 1T$'$/2H WSe$_2$ heterobilayer and its spin-orbit-induced band gap and layer projection. 
    (a) Side (top) and top-down (bottom) views of the $Mh$ stacking configuration (W, grey; Se, green).
    (b) Band structure without SOC. The system is semi-metallic with a band inversion of valence and conduction bands along the $\Gamma$-X direction.
    (c) Same calculation including SOC. Colors denote the out-of-plane spin expectation value. SOC opens a direct gap of 73~meV and lifts spin degeneracies away from time-reversal invariant momenta.
    (d) Layer-resolved projection of the SOC band structure. Blue, red, and green correspond to electronic weight localized in the 1T$'$ layer, the 2H layer, and an equal hybrid of both, respectively. States near the Fermi level arise almost exclusively from the 1T$'$ layer, while the 2H valence- and conduction-band edges reside $\sim$0.3~eV below and $\sim$1.2~eV above E$_\mathrm{F}$.}
    \label{fig:ab1}
\end{figure}

\begin{figure}[H]
    \centering
    \includegraphics[width=\textwidth]{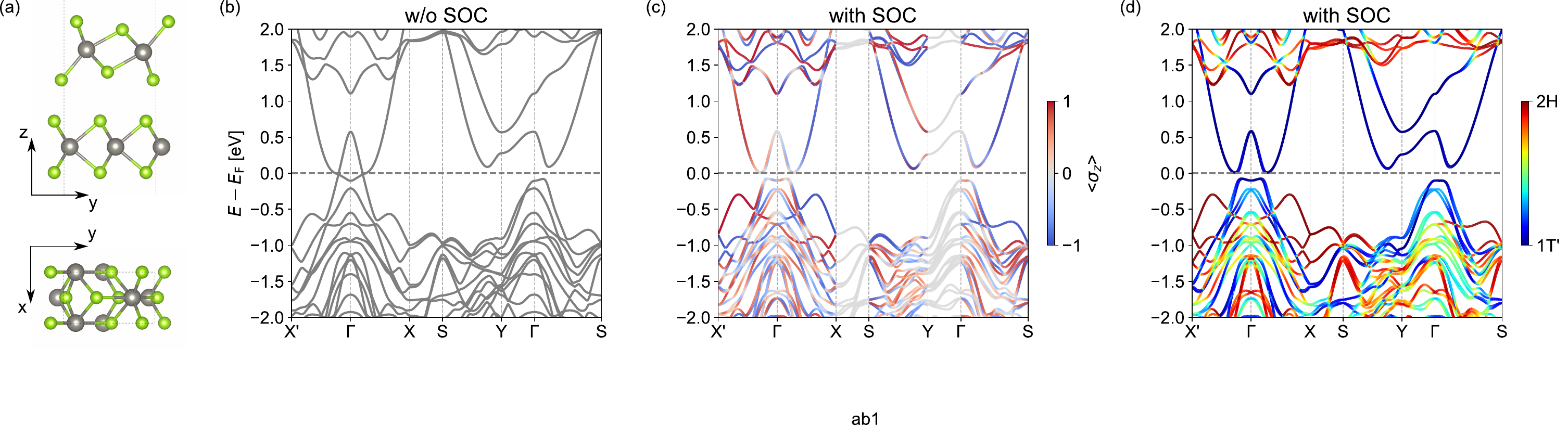}
    \caption{Atomic structure of a $Xh$ stacking 1T$'$/2H WSe$_2$ heterobilayer and its spin-orbit-induced band gap and layer projection in a commensurate 1T$'$/2H WSe$_2$ heterobilayer. 
    (a) Side (top) and top-down (bottom) views of the $Xh$ stacking configuration (W, grey; Se, green).
    (b) Band structure without SOC. The system is semi-metallic with a band inversion of valence and conduction bands along the $\Gamma$-X direction.
    (c) Same calculation including SOC. Colors denote the out-of-plane spin expectation value. SOC opens a direct gap of 78~meV and lifts spin degeneracies away from time-reversal invariant momenta.
    (d) Layer-resolved projection of the SOC band structure. Blue, red, and green correspond to electronic weight localized in the 1T$'$ layer, the 2H layer, and an equal hybrid of both, respectively. States near the Fermi level arise almost exclusively from the 1T$'$ layer, while the 2H valence- and conduction-band edges reside $\sim$0.3~eV below and $\sim$1.2~eV above E$_\mathrm{F}$.}
    \label{fig:a1a}
\end{figure}

Since the 1T$'$ phase can be considered a distortion of the hexagonal lattice, the free-standing monolayers exhibit a similar lattice parameter in one direction: the 2H phase has a hexagonal lattice constant of 3.297 \AA, and the 1T$'$ phase has lattice parameters of a = 3.299 \AA\ and b = 5.997 \AA\ (Table \ref{tab:lattice_hs}).. The resulting heterostructure contains 12 atoms -- 6 from each layer -- arranged in a rectangular supercell. Structure relaxations from various initial alignments converge to two distinct and stable configurations both with optimized lattice parameters of approximately a = 3.27 \AA and b = 5.86 \AA. We label these two nonequivalent stackings as "Se-centered" (labeled ($Xh$)) and "W-centered" (labeled ($Mh$))(see Figure \ref{fig:ab1} (a) and Figure \ref{fig:a1a} (a)). These configurations differ by a 180\(^{\circ}\) in-plane rotation of the 2H layer, which alters the interlayer atomic alignment. Within the 2H layer, W atoms form a triangular sublattice with hexagonal holes at the centers. The key difference between the two stackings is which atoms of the 1T$'$ layer align with these hexagonal holes:
in the $Mh$ stacking, the hexagonal holes align with the W atoms of the 1T$'$ layer, particularly with the W-W zigzag chains. Conversely, in the $Xh$ stacking, the hexagonal holes of the 2H layer are positioned above the Se atoms of the 1T$'$ layer. Both stackings preserve the two-fold rotation axis (C\(_{2}\)) perpendicular to the layers.

Without SOC, the heterostructure is semi-metallic and exhibits a band inversion near the $\Gamma$ point (Figure~\ref{fig:ab1}(a)), inherited from the 1T$'$ WSe$_2$ monolayer as presented above. 
As shown in Figure~\ref{fig:ab1}(b), the inclusion of SOC opens a direct gap of 73~meV at the band inversion point. This gap is notably smaller than the 106~meV gap of the free-standing 1T$'$ monolayer, due to the strain applied to form the commensurate supercell (see Table~\ref{tab:lattice_hs}).
Furthermore, SOC lifts spin degeneracy along the $\Gamma$-X direction, introducing such out-of-plane spin polarizations in Figure~\ref{fig:ab1}(b), making the heterostructure as an attractive candidate for spin-valley optoelectronics.

To quantify the interlayer hybridization, we calculated the layer-projected SOC band structure, shown in Figure~\ref{fig:ab1}(c). Near the Fermi level, the electronic states are almost exclusively localized on the 1T$'$ layer (blue). The 2H monolayer's VBM lies $\sim$0.3~eV below E$_\mathrm{F}$, while its CBM sits $\sim$1.2~eV above. Significant hybridization (green) emerges only at higher energies, indicating interlayer electronic coupling near the band edges is weak. Importantly, the band-edge states that define the topological gap show no contribution from the 2H layer, suggesting the QSH states spatially reside on the 1T$'$ layer, despite the van der Waals interaction with the 2H layer. 
These findings demonstrate that the 1T$'$/2H-WSe$_2$ heterostructure forms a conventional type-I heterostructure where the band edges reside in the 1T$'$ layer, and the 2H layer functions as an electronically passive, wide-band-gap substrate.

Contrasting $Mh$ and $Xh$ stacking presents a geometric control for spin texture, and thus a possibility to influence the interaction between excitons and topologically protected edge states. Because the 2H monolayer lacks inversion symmetry, its valley-contrasting spin splitting reverses under a 180° in-plane rotation that exchanges K and K' valleys. The color scale in Figure~\ref{fig:a1a} (b) flips for every 2H-dominated state compared to Figure~\ref{fig:ab1} (c). This effect is more clearly observed at the VBM of the 2H layer, which flips from predominantly red (spin-up) in $Xh$ to a blue (spin-down) in $Mh$. The small concomitant change in the gap (5 meV) is attributed to relaxation-induced strain rather than electronic hybridization.

\subsection{1T$'$/2H stacking: moiré superlattices}

\begin{figure}[htbp]
    \centering
    \includegraphics[width=0.48\textwidth]{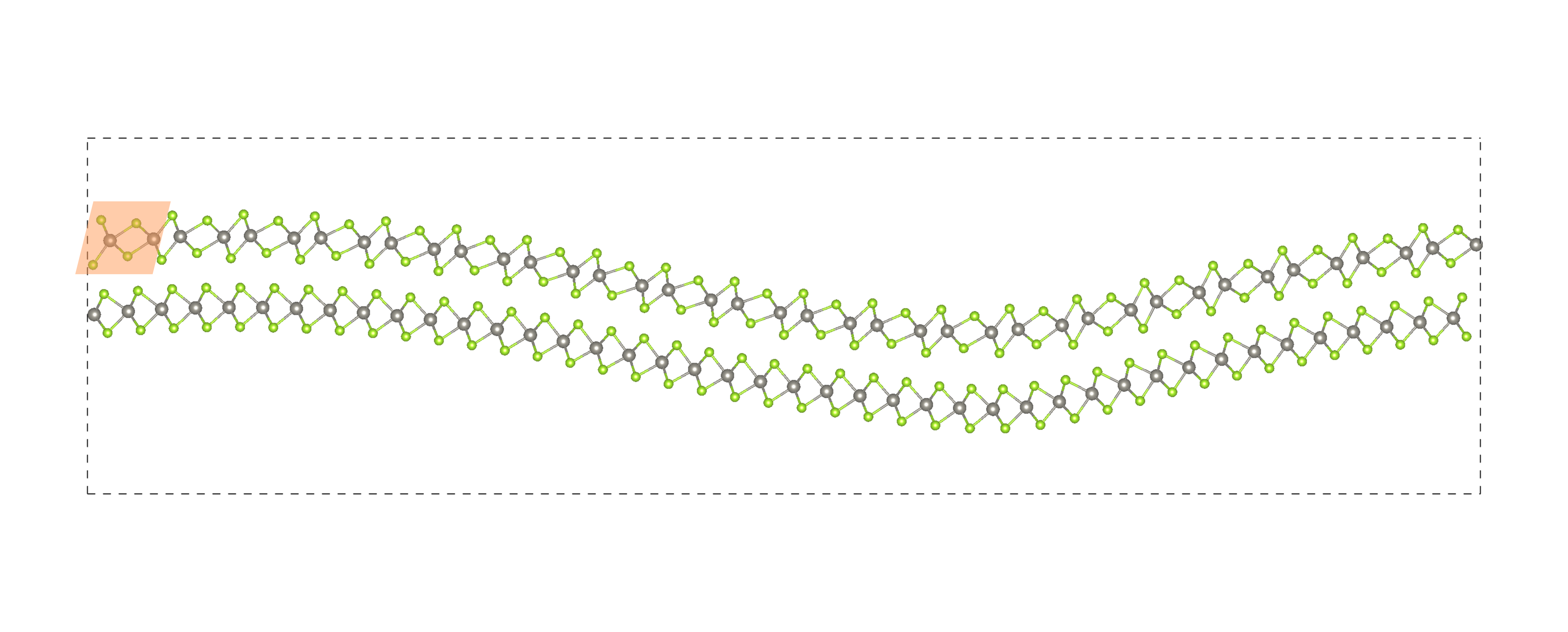}
    \caption{Side view of the fully relaxed atomic structure of the relaxed 1T$'$/2H-WSe$_2$ heterostructure at a 0$^{\circ}$ twist angle. The dashed box indicates the supercell boundaries. W atoms are grey, Se are light green. The orange shaded region highlights the six-atom primitive cell of the 1T$'$ layer used for the band projection analysis in Figure~\ref{fig:0deg}d.}
    \label{fig:moire-0deg}
\end{figure}

\begin{figure}[htbp]
    \centering
    \includegraphics[width=\textwidth]{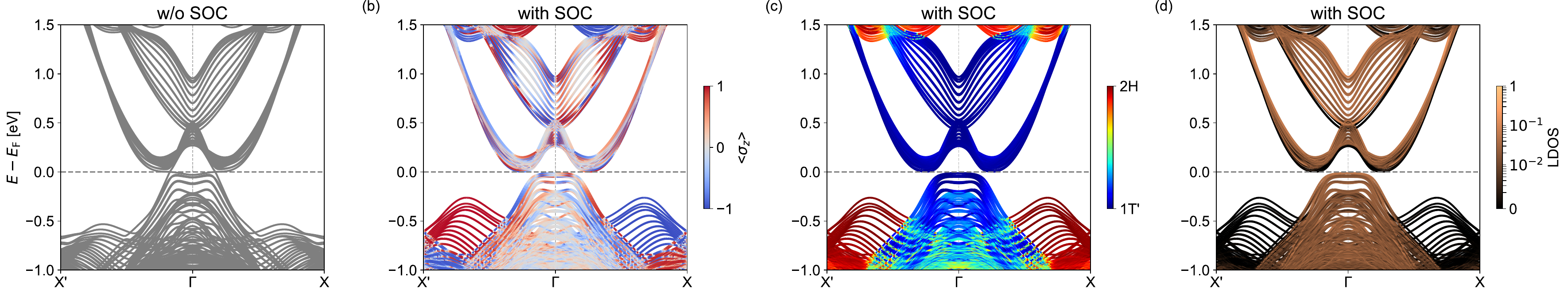}
    \caption{Miniband structure of the relaxed 1T$'$/2H-WSe$_2$ heterobilayer at 0$^{\circ}$. (a) Electronic dispersion without SOC along X'-$\Gamma$-X of the mini-Brillouin zone, showing a metallic state.
    (b) Same calculation but with SOC included. SOC opens a gap of $\approx$15~meV.
    (c) SOC band structure colored by layer weight.
    (d) Projection of the SOC bands onto the primitive cell of the 1T$'$ layer (highlighted in Figure~\ref{fig:moire-0deg}). Only part of the minibands near the Fermi level preserve LDOS, revealing that original bands of the monolayer are broken into a series of flattened, spatially separated minibands -- electronic localization within the moiré superlattice}
    \label{fig:0deg}
\end{figure}

Moving beyond the idealized commensurate heterostructure, we now turn to more realistic models that include the effects of lattice mismatch and lattice reconstruction. 
We begin by generating a moiré superlattice with a twist angle of 0°.
The moiré pattern arises from the small lattice mismatch between the 1T$'$ and 2H phases and is generated by combining $1\times20$ 1T$'$ and $1\times21$ 2H-orthogonal unit cells (see Table \ref{tab:lattice_moire}), which were then fully relaxed (Figure~\ref{fig:moire-0deg}).
Significant atomic reconstruction creates a strong and spatially inhomogeneous strain field which alters the electronic properties.
The resulting bands, folded into the mini-Brillouin zone, are dramatically reconstructed (Figure~\ref{fig:0deg}). 
In the absence of SOC (Figure~\ref{fig:0deg}(a)), the system remains metallic owing to band crossings inherited from the semi-metallic 1T$'$ layer.
Upon including SOC, a gap of only 15~meV opens(Figure~\ref{fig:0deg}(b)), much lower than the 106~meV in the free-standing 1T$'$ monolayer.
The layer-resolved projection in Figure~\ref{fig:0deg}(c) confirms that the type-I band alignment is preserved, with the low-energy states remaining localized on the 1T$'$ layer. 
This suggests that the dramatic band gap reduction is not caused by direct interlayer hybridization, but by the lattice reconstruction (corrugation) of the 1T$'$ layer (see Figure~\ref{fig:moire-0deg}).
To better understand this band reconstruction, we project the superlattice minibands onto the primitive six-atom cell of the 1T$'$ layer (highlighted in Figure~\ref{fig:moire-0deg}), with the results shown in Figure~\ref{fig:0deg}d.
For an ideal, unstrained 1T$'$ monolayer, such a projection onto its primitive cell would faithfully reproduce the original, continuous band dispersion. Here, however, the projection (Figure~\ref{fig:0deg}d) shows that only part of the minibands near the Fermi level carry spectral weight, while the remaining minibands carry negligible weight. This reveals that the primitive bands of the monolayer are broken into a series of flattened sub-bands separated by multiple energy gaps, and electronic states are localized in distinct regions of the moiré supercell.

\begin{figure}[H]
    \centering
    \includegraphics[width=\textwidth]{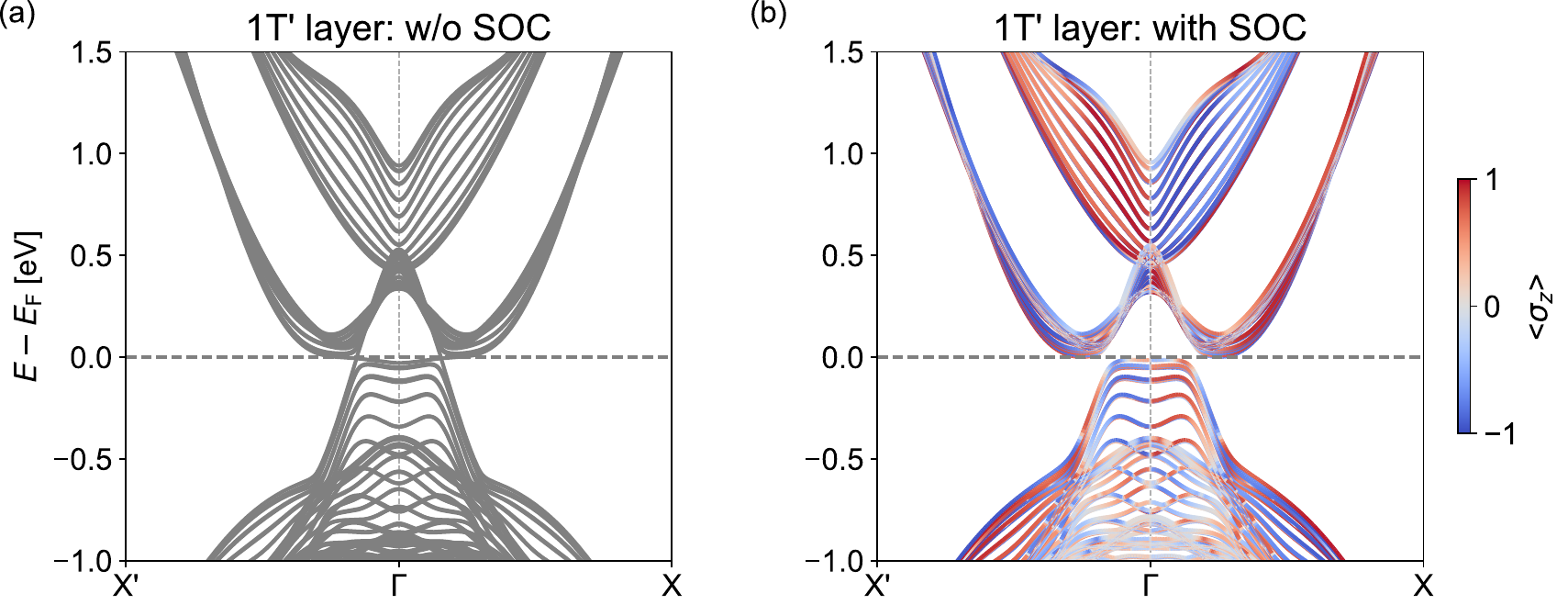}
    \caption{Comparison calculation of an isolated 1T$'$ monolayer using the strained coordinates from the heterostructure at 0$^{\circ}$: (a) with SOC showing a band gap of 0.014~eV, and (b) without SOC showing a metallic state.}
    \label{fig:1t1-0deg}
\end{figure}

To distinguish the effects of the structural strain from direct interlayer electronic hybridization, we performed a comparison calculation on an isolated 1T$'$ monolayer deformed with the same atomic positions as in the relaxed heterostructure (Figure~\ref{fig:1t1-0deg}). 
The calculation on the isolated, corrugated monolayer gives a band gap of 14~meV, nearly identical to the 15~meV gap of the full heterostructure. This result confirms that the electronic reconstruction in the 0$^{\circ}$ moiré superlattice is driven not by interlayer hybridization, but rather by the periodic strain field that the 2H substrate mechanically imparts on the 1T$'$ layer. The 2H layer, therefore, acts as an electronically passive substrate that induces a periodic strain field in the active 1T$'$ layer.

\begin{figure}[htbp]
    \centering
    \includegraphics[width=0.48\textwidth]{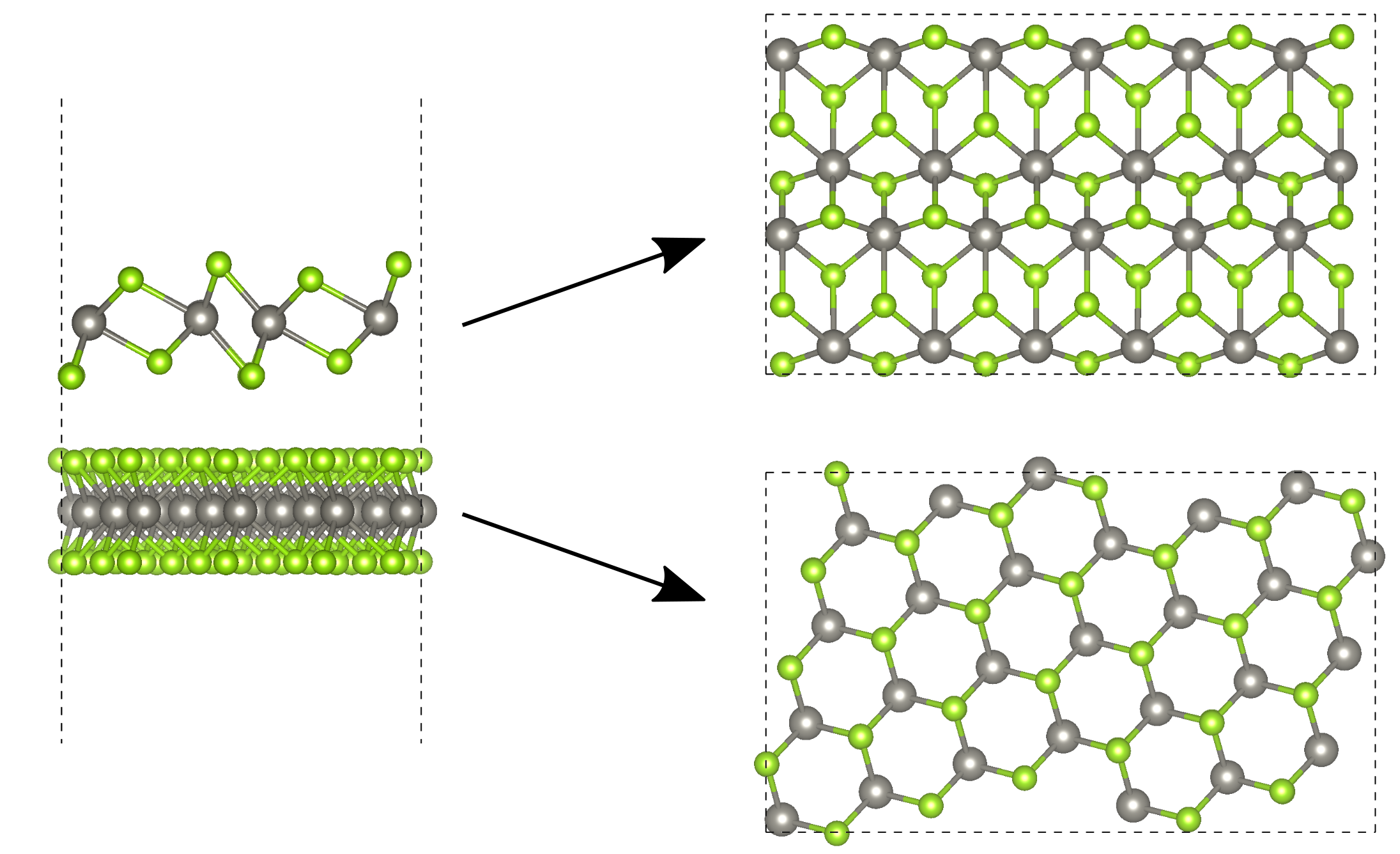}
    \caption{Atomic structure of the relaxed 1T$'$/2H-WSe$_2$ heterostructure at 16.10$^{\circ}$. Left: side view. Right: top-down views of the 1T$'$ (top) and 2H (bottom) layers, which show minimal reconstruction. W atoms are grey, Se are light green. The dashed boxes indicate the supercell boundaries. A maximum of 0.9\% strain is applied to generate the initial geometry.}
    \label{fig:moire-16deg}
\end{figure}

\begin{figure}[htbp]
    \centering
    \includegraphics[width=\textwidth]{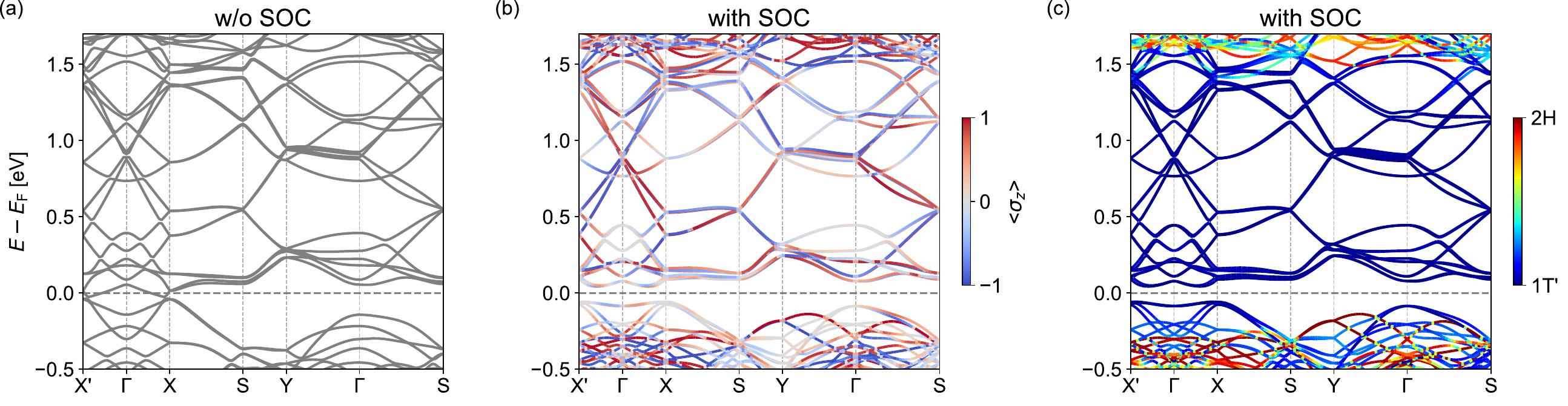}
    \caption{Miniband structure of the relaxed 1T$'$/2H-WSe$_2$ heterobilayer at 16.10$^{\circ}$. (a) Electronic dispersion without SOC along the X'-$\Gamma$-X-S-Y-$\Gamma$-S path of the mini-Brillouin zone, showing a metallic state.
    (b) Same calculation but with SOC included, which opens a gap of 103~meV.
    (c) SOC band structure colored by layer weight.}
    \label{fig:16deg}
\end{figure}

\begin{figure}[H]
    \centering
    \includegraphics[width=\textwidth]{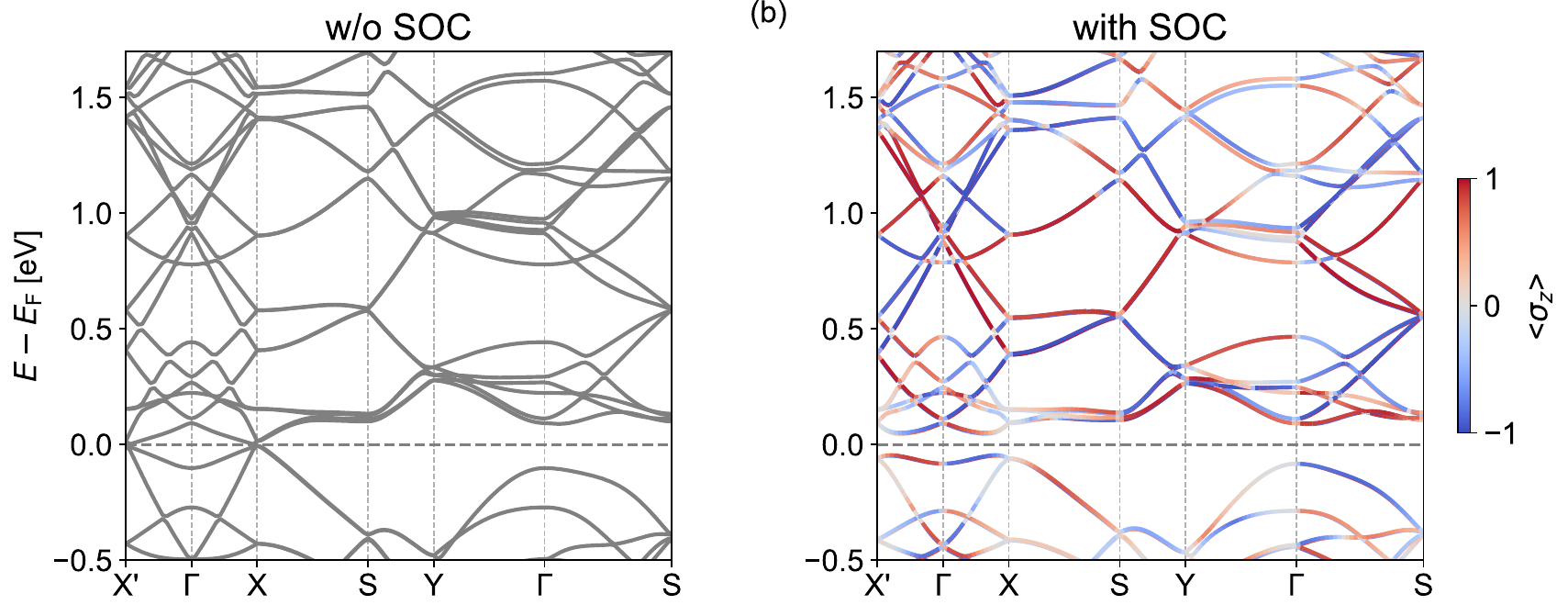}
    \caption{Comparison calculation of an isolated 1T$'$ monolayer using the strained coordinates from the heterostructure at 16.10$^{\circ}$: (a) with SOC showing a band gap of 0.108~eV, and (b) without SOC showing a metallic state.}
    \label{fig:1t1-16deg}
\end{figure}

Finally, to explore the opposite limit of weak interlayer coupling, we investigated a heterostructure with a large twist angle -- 16.10$^{\circ}$. In contrast to the 0$^{\circ}$ case, the relaxed structure shows negligible atomic reconstruction (Figure~\ref{fig:moire-16deg}). 
This mechanical decoupling translates directly into electronic decoupling.
Figure~\ref{fig:16deg} shows the calculated electronic dispersion for the 16.10$^{\circ}$ twisted heterobilayer.
Without SOC (Figure~\ref{fig:16deg}(a)), the system is semi-metallic, consistent with the results for both the free-standing 1T$'$ monolayer and the high-symmetry heterostructure. With SOC included (Figure~\ref{fig:16deg}(b)), the band structure preserves the dispersive character of the 1T$'$ monolayer, with E$_g$ = 103~meV. This value is nearly identical to the gaps obtained for a free-standing 1T$'$ monolayer (106~meV) and importantly, also to the gap of an isolated 1T$'$ layer subjected to the same uniform strain used to construct the 16.10$^{\circ}$ superlattice (108~meV, see Figure~\ref{fig:1t1-16deg}). Layer-projected analysis (Figure~\ref{fig:16deg}(c)) confirms this electronic decoupling, showing that the low-energy electronic states are completely localized in the 1T$'$ layer.
These results demonstrate that the large twist angle effectively isolates the 1T$'$ layer, switching off strain effects dominant at the small angle and direct electronic hybridization, creating a system that preserves the intrinsic QSH character of the 1T$'$ monolayer.

\subsection{Edge states}

The previous sections have shown that a 1T$'$/2H bilayer can be tuned from electronically decoupled (at the large twist angle) to reconstructed (at the small twist angle), without closing the bulk quantum-spin-Hall gap of the 1T$'$ monolayer. It remains unresolved whether the resulting heterostructure still hosts the helical edge channels that characterize the 1T$'$ phase as a 2D TI.

We start with 1T$'$ monolayer as reference. Then to disentangle the contributions from 1T$'$ and 2H, we construct two complementary nanoribbon systems. In the first, a finite-width 1T$'$ ribbon is placed on an extended 2H monolayer without edges in the simulation cell, simulating a large 2H substrate that cannot contribute edge states. In the second, we create a heteroribbon with a 2H ribbon which is slightly wider than the 1T$'$ ribbon, representing finite flakes and allowing us to quantify the energetic proximity and spatial leakage of any 2H-edge-derived states to the 1T$'$ edges. For all geometries we choose zigzag terminations where the SOC-induced topological gap is preserved, and we use fully relaxed geometries

\begin{figure}[htbp]
    \centering
    \includegraphics[width=\textwidth]{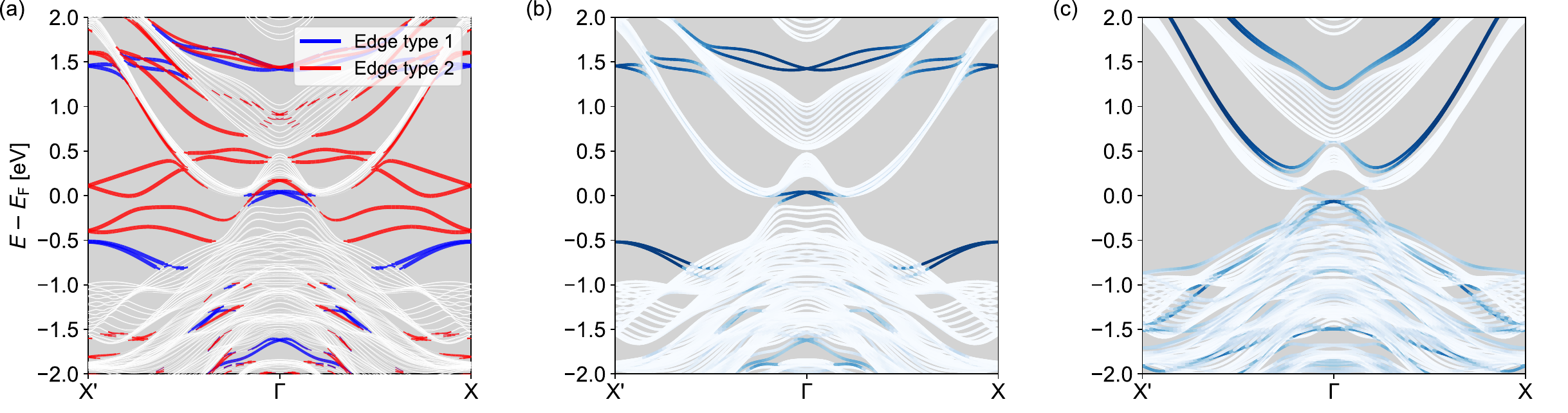}
    \caption{Robust quantum-spin-Hall edge states in monolayer 1T$'$-WSe$_2$.
    (a) Edge-projected electronic band structure for a ribbon hosting type 1 (blue) and type 2 (red) terminations on opposite sides. White shading marks the projected bulk bands. Both edges exhibit a linearly dispersing, gapless state that bridges the inverted bulk gap at $\Gamma$.
    (b) Band structure for a symmetric ribbon with edge type 1 on both sides. Two degenerate Dirac edge modes appear (blue).
    (c) Same geometry as (b) after full hydrogen passivation. The in-gap crossing is preserved. Energies are referenced to the Fermi level. All calculations include spin-orbit coupling and use fully relaxed geometries.}
    \label{fig:ed-mono}
\end{figure}

\begin{figure}[H]
    \centering
    \includegraphics[width=\textwidth]{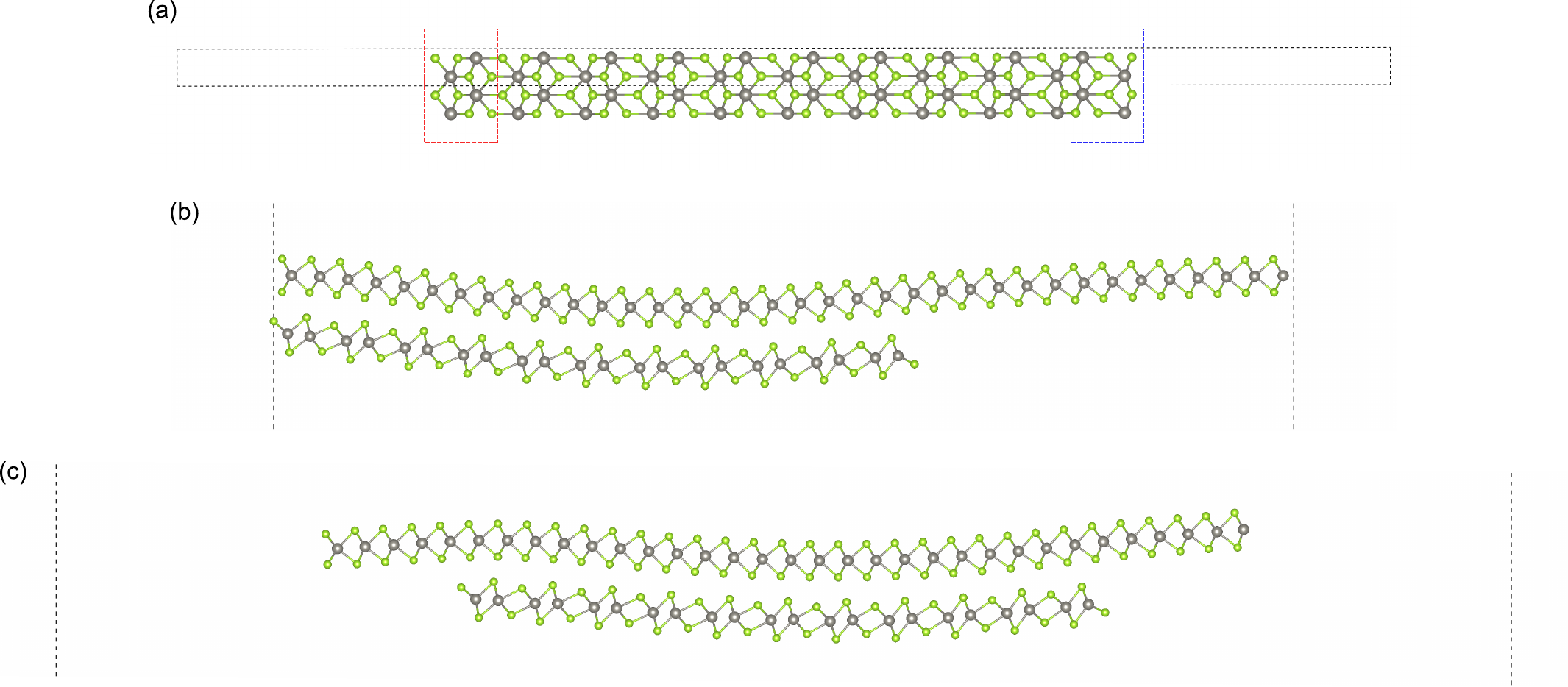}
    \caption{(a) Atomic structure of a 1T$'$-WSe$_2$ nanoribbon. The left (red box) and right (blue box) terminations define edge types 1 and 2, respectively. Edge type 1 cuts the longer W--Se bonds; edge type 2 terminates on the shorter bonds. As cleavage along the longer bonds is expected to align with experimentally accessible exfoliation directions, type 1 edge is used for bilayer simulations.
    (b) A finite-width 1T$'$ ribbon placed on an extended, edgeless 2H-WSe$_2$ monolayer.
    (c) Heteroribbon in which the 2H ribbon is slightly wider than the 1T$'$ ribbon.
     Dashed lines denote the periodic supercell}
    \label{fig:ed-structure}
\end{figure}

Figure~\ref{fig:ed-mono} demonstrates the robustness of QSH edge modes in monolayer 1T$'$-WSe$_2$. Edge-projected band calculations show that both crystallographically distinct terminations (see Figure \ref{fig:ed-structure} (a)) host gapless, linearly dispersing states that connect the inverted bulk gap and cross at the Fermi level, a character of 2D TI.
We define two distinct edge terminations: type 1, created by severing the longer W--Se bonds, and type 2, from cleaving the shorter W--Se bonds. As cleavage along the longer bonds is expected to align with experimentally accessible exfoliation directions, we focus on the type 1 edge for the calculations presented in panels (b), (c), and all subsequent bilayer simulations.
For a symmetric ribbon with type 1 terminations on both sides (panel (b)), the helical edge modes localized on opposite edges persist, appearing as two nearly degenerate Dirac crossings in the band structure.
Upon hydrogen passivation (panel (c)), the in-gap linear crossing is preserved, whereas termination-induced dangling-bond features are removed or pushed into the bulk continua.

\begin{figure}[htbp]
    \centering
    \includegraphics[width=\textwidth]{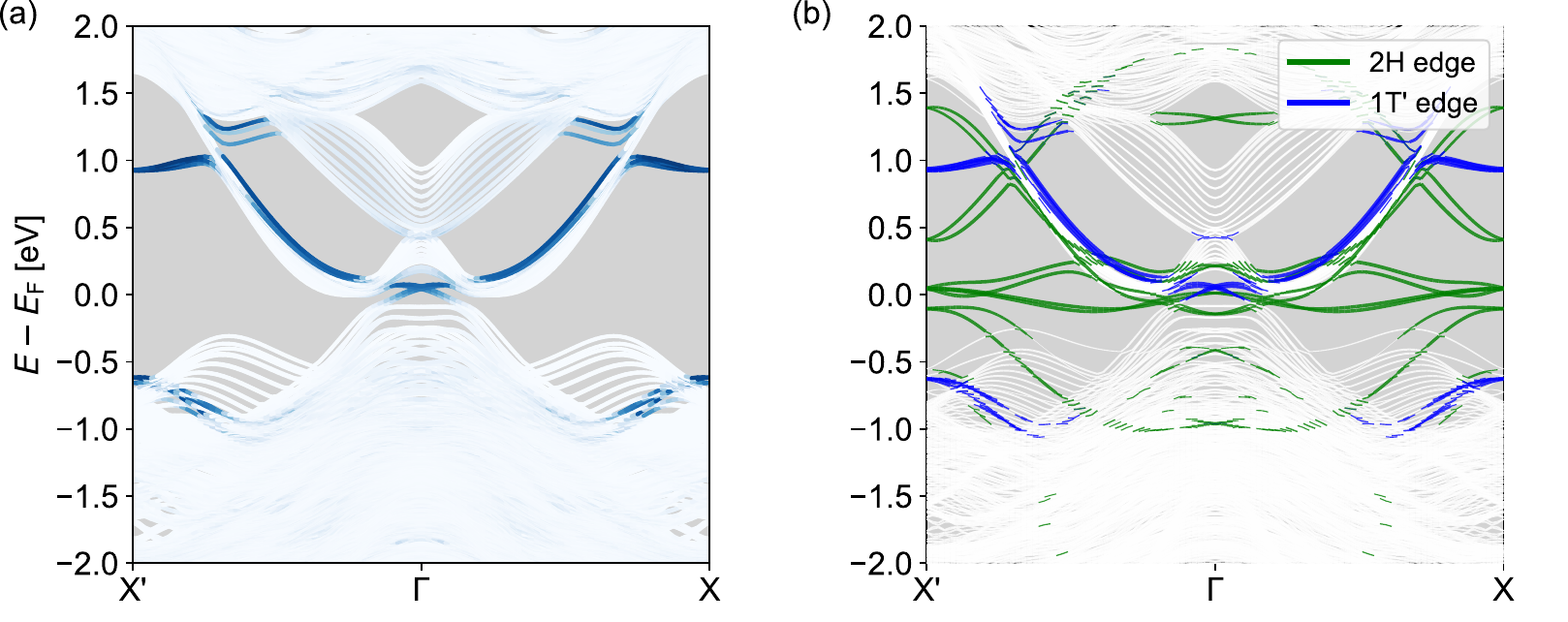}
    \caption{Persistence of the 1T$'$ helical edge modes and emergence of 2H-derived states in realistic heterobilayers.
    (a) Same 1T$'$ ribbon as in Figure~\ref{fig:ed-mono}(b) placed on a laterally infinite, edgeless 2H-WSe$_2$ monolayer that simulates a macroscopic substrate. After full structural relaxation, the helical branch is preserved with a small upward energy shift.
    (b) Heteroribbon consisting of a slightly wider 2H ribbon stacked under the 1T$'$ ribbon, so that both layers expose edges. The topological Dirac branch (blue) remains, whereas the 2H edges generate a dense set of bands (green) within the nominal 1T$'$ gap. White shading denotes the total projected bulk continuum. All calculations include spin--orbit coupling and use fully relaxed geometries. The absence of avoided crossings confirms negligible direct hybridization between the two layers.}
    \label{fig:ed-bi}
\end{figure}

Having established the intrinsic edge-state dispersion in isolated ribbons, Figure \ref{fig:ed-bi} presents how a 2H substrate modifies this spectrum.
Figure~\ref{fig:ed-mono}(b) provides a reference, showing that an isolated 1T$'$ ribbon with type 1 edges hosts a pair of degenerate helical modes, which would form a single idealized Dirac crossing at $\Gamma$ in the semi-infinite limit. Introducing a van der Waals substrate, modeled as a laterally infinite 2H monolayer (see Figure \ref{fig:ed-structure} (b)), does not fundamentally alter the 1T$'$ helical spectrum, as shown in Figure~\ref{fig:ed-bi}(a). This indicates that there is no direct electronic interlayer hybridization that would gap or strongly renormalize the edge states. The upward shift of the Dirac point is attributed to the long-wavelength corrugation that develops upon relaxation, but no additional states appear inside the bulk gap. This observation confirms our earlier conclusion that direct electronic hybridization between 1T$'$ and 2H layers is negligible; an edgeless 2H substrate remains spectroscopically passive around $E_F$.

The results change fundamentally once the 2H layer itself is terminated. Heteroribbons (see Figure \ref{fig:ed-structure} (c)), representative of realistic finite flakes or nanostructures, introduce a series of Se- and W-terminated zigzag edges whose dispersion branches (green) appear in the same energy window as the 1T$'$ helical modes (blue), as shown in Figure \ref{fig:ed-bi} (b). The absence of pronounced anticrossings at near-degeneracies indicates weak direct hybridization between the 1T$'$ and 2H edge channels, consistent with their lateral offset and interlayer spacing.
From an experimental standpoint these findings have two immediate implications. First, topological transport signatures (quantized conductance plateaus, non-local edge currents) remain theoretically accessible even in the presence of a 2H substrate, provided that the 2H layer is macroscopic or otherwise edge-free in the device active region. Second, any scanning-probe or photoemission experiment performed on finite 1T$'$/2H heteroflakes must disentangle the desired helical branch from a dense set of trivial 2H edge resonances that lie spectrally nearby, otherwise these trivial bands may be misinterpreted as 1T$'$ edge modes.

\section{Conclusion}

We implemented a self-consistent spin-orbit-coupling scheme within the DFTB and GFN1-xTB framework in the AMS suite. The accuracy of both implementations was validated against established benchmark calculations, as detailed in the Supporting Information.
This method enables computationally efficient simulations of SOC-driven phenomena in large-scale systems, and we applied it to 1T$'$-WSe$_2$ monolayers and their 1T$'$/2H-WSe$_2$ heterostructures.

Across all investigated twist angles, the active electronic states near the Fermi level predominantly originate from the 1T$'$ layer. The 2H layer's band-edge states lie significantly higher or lower in energy, making it electronically passive and thus primarily acting as a substrate. 
At the small twist angle, the lattice relaxation reconstructs the minibands of the 1T$'$ layer, arising from the structural deformation of the 1T$'$ layer itself, rather than from direct interlayer electronic hybridization. A direct comparison shows that an isolated 1T$'$-WSe$_2$ monolayer subjected to the same strain and lattice distortion as in the heterostructure exhibits an almost identical band gap and band edges.
At the large twist angle, the two layers become electronically decoupled, and the system preserves its 1T$'$ layer's characteristics, though subject to strain effects.

At the edges, monolayer 1T$'$ hosts robust QSH helical modes. 
An edge-free, laterally infinite 2H substrate preserves these 1T$'$ edge modes, inducing only a shift of the Dirac point due to structural relaxation and introducing no additional in-gap states. When the 2H layer is terminated, its zigzag edges contribute trivial dispersion branches in the same energy window with weak hybridization between topological and trivial channels, so intrinsic helical transport can persist but may be spectroscopically obscured.

In summary, our findings reveal that structural relaxation dominates moiré miniband formation, interlayer electronic hybridization is negligible near the Fermi level, and edge physics is robust to an edgeless substrate yet sensitive to substrate edges. These results provide practical guidance, suggesting that quantized conductance plateaus and non-local edge currents should be observable on macroscopic, edge-free 2H supports, whereas scanning-probe and photoemission studies of finite heteroflakes must disentangle helical modes from nearby substrate-edge resonances. Beyond WSe$_2$, the validated methodology presented in this work offers a computationally efficient and predictive tool for exploring SOC-enabled phases and devices at experimentally relevant sizes.

\section*{Acknowledgements}
Financial support of the European Commission via Marie S. Curie ITN 2Exciting (GA 956813) and the Deutsche Forschungsgemeinschaft via the Collaborative Research Center 1415 (Chemistry of Synthetic 2D Materials) and the German-Israelian project initiative HE 3543/42-1 is gratefully acknowledged. The authors acknowledge the computing time made available to them on the high-performance computer at the NHR Center of TU Dresden. This center is jointly supported by the Federal Ministry of Education and Research and the state governments participating in the NHR (www.nhr-verein.de/unsere-partner).
The authors would like to thank Software for Chemistry \& Materials B.V. (SCM) for their support during the code implementation, in particular Dr. Erik van Lenthe, Dr. Matti Hellström, Dr. Robert Rüger, and Dr. Bas Rustenburg. The authors are grateful to Gautam Jha for insightful discussions and for sharing his expertise in spin-orbit coupling and DFTB.

\section*{Contributions}
T. Brumme and T. Heine conceptualized and supervised the research. W. Li implemented the computational code with support from P. Philipsen, performed the calculations, and analyzed the data. All authors contributed to the discussion and manuscript writing.

\newpage
\section*{SI}
\subsection*{Additional data for 1T$'$/2H-WSe$_2$ heterobilayer}

\begin{table}[H]
    \centering
    \begin{tabular}{lcc}
        \hline
        Structure & $a$ (\AA) & $b$ (\AA) \\
        \hline
        1T$'$ & 3.299 & 5.997 \\
        2H & 3.297 & - \\
        HS-$Xh$ & 3.271 & 5.860 \\
        HS-$Mh$ & 3.269 & 5.860 \\
        \hline
    \end{tabular}
    \caption{Lattice constants for relaxed free-standing monolayers and the high-symmetry commensurate heterostructure (HS). The $a$ and $b$ lattice vectors correspond to the zigzag and armchair directions. The strain applied to generate HS is nearly identical in both $Xh$ and $Mh$ configurations: the 1T$'$ layer by -0.9\% (zigzag) and -2.3\% (armchair) and the 2H layer by -0.8\% (zigzag) and +2.6\% (armchair).}
    \label{tab:lattice_hs}
\end{table}

\begin{table}[H]
    \centering
    \begin{tabular}{lcc}
        \hline
        Structure & $a$ (\AA) & $b$ (\AA) \\
        \hline
        Relaxed  & 3.269 & 117.410 \\
        Rigid & 3.297 & 119.936 \\
        \hline
    \end{tabular}
    \caption{Lattice constants for the 0$^{\circ}$ twist angle moiré superlattice before (Rigid) and after (Relaxed) structural optimization. The $a$ and $b$ lattice vectors correspond to the zigzag and armchair directions, respectively. The superlattice was constructed from a 1x20 1T$'$ supercell and a 1x21 2H-orthogonal supercell, corresponding to a initial lattice mismatch of $\sim$0.05\% (zigzag) and $\sim$0.02\% (armchair).}
    \label{tab:lattice_moire}
\end{table}

\subsection*{Benchmark calculations}

The validation of our spin-orbit coupling (SOC) implementation in the AMS/DFTB software was conducted by performing benchmark calculations on a series of molecules and 2D materials, comparing our results against those obtained from DFTB+ software and available experimental data. The analysis focused on key parameters such as the HOMO-LUMO gap, SOC energy, bond length, and vibrational frequencies, band splitting energies.

\subsubsection*{Close shell molecules}

For the HOMO-LUMO gap and spin-orbit coupling energy benchmarks (Tables \ref{tab:HLgap} and \ref{tab:SOenergy}), calculations on the diatomic molecules were performed at their experimental bond lengths~\cite{Huber1979}. For the bond length and vibrational frequency benchmarks (Tables \ref{tab:Bond_Length} and \ref{tab:Frequency}), the geometries were optimized, and the results are compared with experimental values.

\begin{table}[H]\setlength{\tabcolsep}{4pt}
    \centering
    \caption{HOMO-LUMO gap (eV) for various molecules calculated using different methods. SO: with spin-orbit coupling; NR: non-relativistic; $\Delta_{\text{SO}}$ : difference between SO and NR calculations. Results are shown for DFTB with QUASINANO13 parameters and GFN1-xTB methods using both AMS/DFTB and DFTB+ software. The 'Diff' row shows the difference between AMS/DFTB and DFTB+ $\Delta_{\text{SO}}$ values, highlighting the consistency between implementations.}
    \begin{tabular}{cccccccccccc}
    \hline\hline
            &       &       & Bi$_2$ & HI    & I$_2$  & IF    & PbO   & PbTe  & TlF   & TlH   & TlI   \\
    \hline
    SO      & \multirow{7}[0]{*}{DFTB2} & \multirow{3}[0]{*}{\shortstack{AMS/\\DFTB}} & 0.835  & 5.389  & 2.461   & 3.718  & 3.495  & 1.704  & 4.796  & 3.425  & 3.184  \\
    NR      &       &       & 1.854  & 5.802  & 2.846   & 4.070  & 4.016  & 2.431  & 5.283  & 3.856  & 3.831  \\
    $\Delta_{\text{SO}}$ &       &       & -1.019 & -0.413 & -0.385  & -0.352 & -0.521 & -0.727 & -0.488 & -0.431 & -0.646 \\
    \cline{3-12}
    SO      &       & \multirow{3}[0]{*}{DFTB$+$}         & 0.835  & 5.389  & 2.461   & 3.718  & 3.495  & 1.704  & 4.796  & 3.425  & 3.184  \\
    NR      &       &       & 1.854  & 5.802  & 2.846   & 4.070  & 4.016  & 2.431  & 5.283  & 3.856  & 3.831  \\
    $\Delta_{\text{SO}}$ &       &       & -1.019 & -0.413 & -0.385  & -0.352 & -0.521 & -0.727 & -0.488 & -0.431 & -0.646 \\
    \cline{3-12}
    \textbf{Diff} &       &       & \textbf{0.000} & \textbf{0.000} & \textbf{0.000} & \textbf{0.000} & \textbf{0.000} & \textbf{0.000} & \textbf{0.000} & \textbf{0.000} & \textbf{0.000} \\
    \hline
    SO      & \multirow{7}[0]{*}{GFN1-xTB} & \multirow{3}[0]{*}{\shortstack{AMS/\\DFTB}} & 0.862  & 5.644  & 2.441   & 2.749  & 3.078  & 1.869  & 3.962  & 3.631  & 3.837  \\
    NR      &       &       & 1.917  & 6.054  & 2.829   & 3.065  & 3.597  & 2.666  & 4.452  & 4.027  & 4.359  \\
    $\Delta_{\text{SO}}$ &       &       & -1.055 & -0.411 & -0.389  & -0.316 & -0.519 & -0.797 & -0.490 & -0.395 & -0.522 \\
    \cline{3-12}
    SO      &       & \multirow{3}[0]{*}{DFTB$+$}         & 0.862  & 5.643  & 2.440   & 2.750  & 3.078  & 1.869  & 3.961  & 3.634  & 3.837  \\
    NR      &       &       & 1.917  & 6.054  & 2.829   & 3.066  & 3.597  & 2.666  & 4.452  & 4.029  & 4.359  \\
    $\Delta_{\text{SO}}$ &       &       & -1.055 & -0.411 & -0.389  & -0.316 & -0.519 & -0.797 & -0.490 & -0.395 & -0.522 \\
    \cline{3-12}
    \textbf{Diff} &       &       & \textbf{0.000} & \textbf{0.000} & \textbf{0.000} & \textbf{0.000} & \textbf{0.000} & \textbf{0.000} & \textbf{0.000} & \textbf{0.000} & \textbf{0.000} \\
    \hline\hline
    \end{tabular}
    \label{tab:HLgap}
\end{table}%

\begin{table}[H]\setlength{\tabcolsep}{3pt}
    \centering
    \caption{Spin-orbit coupling energies ($E_{\text{SO}}$) in eV for various molecules. Results are presented for DFTB/QUASINANO13 and GFN1-xTB methods using both AMS/DFTB and DFTB+ software. The 'Diff' row shows the difference between AMS/DFTB and DFTB+ $E_{\text{SO}}$ values, demonstrating the agreement between implementations.}
    \begin{tabular}{cccccccccccc}
    \hline\hline
            &       &       & Bi$_2$ & HI    & I$_2$   & IF    & PbO   & PbTe  & TlF   & TlH   & TlI   \\
    \hline
    $E_{\text{SO}}$     & \multirow{3}[0]{*}{DFTB2} & AMS/DFTB & -2.743 & -0.122 & -0.330 & -0.206 & -0.320 & -0.871 & -0.056 & -0.131 & -0.297 \\
    $E_{\text{SO}}$     &                               & DFTB$+$               & -2.743 & -0.122 & -0.330 & -0.206 & -0.320 & -0.871 & -0.056 & -0.131 & -0.297 \\
    \cline{3-12}
    \textbf{Diff} &                       &                        & \textbf{0.000} & \textbf{0.000} & \textbf{0.000} & \textbf{0.000} & \textbf{0.000} & \textbf{0.000} & \textbf{0.000} & \textbf{0.000} & \textbf{0.000} \\
    \hline
    $E_{\text{SO}}$     & \multirow{3}[0]{*}{GFN1-xTB}   & AMS/DFTB & -3.316 & -0.140 & -0.379 & -0.299 & -0.305 & -0.902 & -0.041 & -0.118 & -0.305 \\
    $E_{\text{SO}}$     &                               & DFTB$+$               & -3.316 & -0.140 & -0.379 & -0.299 & -0.305 & -0.902 & -0.041 & -0.118 & -0.305 \\
    \cline{3-12}
    \textbf{Diff} &                       &                        & \textbf{0.000} & \textbf{0.000} & \textbf{0.000} & \textbf{0.000} & \textbf{0.000} & \textbf{0.000} & \textbf{0.000} & \textbf{0.000} & \textbf{0.000} \\
    \hline\hline
    \end{tabular}
    \label{tab:SOenergy}
\end{table}

\begin{table}[H]\setlength{\tabcolsep}{4pt}
    \centering
    \caption{Bond lengths (\AA) for various molecules. Experimental values~\cite{Huber1979} are provided for reference. SO: with spin-orbit coupling; NR: non-relativistic; $\Delta_{\text{SO}}$: difference between SO and NR calculations. Results are shown for the GFN1-xTB method using both AMS/DFTB and DFTB+ software. Additional $\Delta_{\text{SO}}$ values from GGA and LDA calculations using AMS/ADF are included for comparison.}
    \begin{tabular}{cccccccccccc}
    \hline\hline
            &       &       & Bi$_2$   & HI    & I$_2$    & IF    & PbO   & PbTe  & TlF   & TlH   & TlI \\
    \hline
    exp   &       &       & 2.661  & 1.609  & 2.667   & 1.910  & 1.922  & 2.595  & 2.084  & 1.870  & 2.814  \\
    \hline
    SO    & \multirow{6}[0]{*}{GFN1-xTB} & \multirow{3}[0]{*}{\shortstack{AMS/\\DFTB}} & 2.720  & 1.614  & 2.705  & 1.920  & 1.886  & 2.654  & 2.244  & 1.983  & 2.922  \\
    NR    &       &       & 2.669  & 1.597  & 2.694  & 1.939  & 1.903  & 2.677  & 2.255  & 2.017  & 2.912  \\
    $\Delta_{\text{SO}}$   &       &       & 0.051 & 0.017 & 0.011 & -0.019 & -0.017 & -0.022 & -0.011 & -0.034 & 0.010 \\
    \cline{3-12}
    SO    &       & \multirow{3}[0]{*}{DFTB+} & 2.720  & 1.614  & 2.705  & 1.920  & 1.886  & 2.654  & 2.244  & 1.983  & 2.922  \\
    NR    &       &       & 2.669  & 1.597  & 2.694  & 1.939  & 1.903  & 2.677  & 2.255  & 2.017  & 2.912  \\
    $\Delta_{\text{SO}}$   &       &       & 0.051 & 0.017 & 0.011 & -0.019 & -0.017 & -0.022 & -0.011 & -0.034 & 0.010 \\
    \hline
    $\Delta_{\text{SO}}$   & GGA   & AMS/ADF & 0.030 & 0.003 & 0.022 & 0.011 & -0.002 & 0.004  & -0.013 & -0.031 & -0.014 \\
    $\Delta_{\text{SO}}$   & LDA   & AMS/ADF & 0.024 & 0.003 & 0.019 & 0.011 & -0.003 & 0.002  & -0.008 & -0.033 & -0.015 \\
    \hline\hline
    \end{tabular}%
    \label{tab:Bond_Length}%
\end{table}%

\begin{table}[H]\setlength{\tabcolsep}{4pt}
    \centering
    \caption{Vibrational frequencies (cm$^{-1}$) for various molecules. Experimental values~\cite{Huber1979} are provided for reference. SO: with spin-orbit coupling; NR: non-relativistic; $\Delta_{\text{SO}}$: percentage difference between SO and NR calculations. Results are shown for the GFN1-xTB method using both AMS/DFTB and DFTB+ software. Additional $\Delta_{\text{SO}}$ values from GGA and LDA calculations using AMS/ADF are included for comparison.}
    \begin{tabular}{cccccccccccc}
    \hline\hline
            &       &       & Bi$_2$ & HI    & I$_2$  & IF    & PbO   & PbTe  & TlF   & TlH   & TlI   \\
    \hline
    exp     &       &       & 173    & 2309  & 215    & 610   & 721   & 212   & 477   & 1391  & 150   \\
    \hline
    SO      & \multirow{6}{*}{GFN1-xTB} & \multirow{3}{*}{\shortstack{AMS/\\DFTB}} & 181 & 2349 & 267 & 601 & 689 & 269 & 317 & 1366 & 172 \\
    NR      &       &       & 220    & 2482  & 277    & 571   & 660   & 254   & 314   & 1336  & 176   \\
    $\Delta_{\text{SO}}$ &   &   & $-18\%$ & $-5\%$ & $-4\%$ & $5\%$ & $4\%$ & $6\%$ & $1\%$ & $2\%$ & $-2\%$ \\
    \cline{3-12}
    SO      &       & \multirow{3}{*}{DFTB$+$} & 180 & 2453 & 266 & 553 & 647 & 239 & 312 & 1325 & 172 \\
    NR      &       &       & 219    & 2483  & 277    & 568   & 661   & 255   & 314   & 1332  & 174   \\
    $\Delta_{\text{SO}}$ &   &   & $-18\%$ & $-1\%$ & $-4\%$ & $-3\%$ & $-2\%$ & $-6\%$ & $-1\%$ & $-1\%$ & $-1\%$ \\
    \hline
    $\Delta_{\text{SO}}$ & GGA & AMS/ADF & $-10\%$ & $-1\%$ & $-6\%$ & $-4\%$ & $-1\%$ & $-4\%$ & $1\%$ & $1\%$ & $0\%$ \\
    $\Delta_{\text{SO}}$ & LDA & AMS/ADF & $-8\%$ & $-1\%$ & $-5\%$ & $-3\%$ & $-1\%$ & $-3\%$ & $1\%$ & $1\%$ & $0\%$ \\
    \hline\hline
    \end{tabular}
    \label{tab:Frequency}
\end{table}

\subsubsection*{2D TMDC}

The geometries for the 2D TMDC monolayers were taken from Ref.~\cite{Jha2022}, which were obtained at the DFT level (PBE functional) with AMS-BAND, in conjugation with scalar relativistic corrections at the ZORA level.

\begin{table}[H]
    \centering
    \caption{Band splitting energies at the valence band maximum (VBM)for TMDC monolayers (MoS$_2$, MoSe$_2$, WS$_2$, and WSe$_2$), calculated using the DFTB method with QUASINANO13 parameters. Results from both AMS/DFTB and DFTB+ software are presented for comparison. All values are in eV and calculated at the direct band gap position (K point) in the Brillouin zone.}
    \begin{tabular}{ccccccc}
    \hline\hline
            &       &       & MoS$_2$    & MoSe$_2$    & WS$_2$    & WSe$_2$   \\ 
    \hline
    $\Delta_{\text{VBM}}$     &  \multirow{2}[0]{*}{DFTB}  & \multirow{1}[0]{*}{AMS/DFTB}   & 0.15  & 0.20  & 0.46  & 0.49  \\
    \cline{3-7}
    $\Delta_{\text{VBM}}$     &   & \multirow{1}[0]{*}{\shortstack{DFTB+}}   & 0.15  & 0.20  & 0.46  & 0.49  \\
    \hline\hline
    \end{tabular}
    \label{tab:splitting-tmdc1}
\end{table}

\begin{table}[H]
    \centering
    \caption{Band splitting energies at the valence band maximum (VBM)for TMDC monolayers (MoS$_2$, MoSe$_2$, WS$_2$, and WSe$_2$), calculated using AMS/DFTB software. Results from both the DFTB method with QUASINANO13 parameters and the GFN1-xTB method are presented to demonstrate the consistency across different computational approaches. All values are in eV and calculated at the direct band gap position (K point) in the Brillouin zone.}
    \begin{tabular}{ccccccc}
    \hline\hline
            &       &       & MoS$_2$    & MoSe$_2$    & WS$_2$    & WSe$_2$   \\ 
    \hline
    $\Delta_{\text{VBM}}$     &  \multirow{1}[0]{*}{DFTB/QUASINANO13}  & \multirow{2}[0]{*}{\shortstack{AMS/\\DFTB}}   & 0.15  & 0.20  & 0.46  & 0.49  \\
    \cline{4-7}\cline{2-2}
    $\Delta_{\text{VBM}}$     & \multirow{1}[0]{*}{GFN1-xTB}   &  & 0.18  & 0.23  & 0.56  & 0.61  \\
    \hline\hline
    \end{tabular}
    \label{tab:splitting-tmdc2}
\end{table}

\bibliographystyle{abbrv}
\bibliography{ref}

\begin{thebibliography}{10}

\bibitem{Angeli2021-zf}
M.~Angeli and A.~H. MacDonald.
\newblock $\gamma$ valley transition metal dichalcogenide moiré bands.
\newblock {\em Proc. Natl. Acad. Sci. U. S. A.}, 118(10), Mar. 2021.
\newblock Publisher: Proceedings of the National Academy of Sciences.

\bibitem{Bannwarth2019}
C.~Bannwarth, S.~Ehlert, and S.~Grimme.
\newblock Gfn2-xtb—an accurate and broadly parametrized self-consistent
  tight-binding quantum chemical method with multipole electrostatics and
  density-dependent dispersion contributions.
\newblock {\em J. Chem. Theory Comput.}, 15(3):1652--1671, 2019.

\bibitem{Bernevig2006}
B.~A. Bernevig and S.-C. Zhang.
\newblock Quantum spin hall effect.
\newblock {\em Phys. Rev. Lett.}, 96:106802, Mar 2006.

\bibitem{Bistritzer2011}
R.~Bistritzer and A.~H. MacDonald.
\newblock Moiré bands in twisted double-layer graphene.
\newblock {\em Proc. Natl. Acad. Sci. U.S.A.}, 108(30):12233--12237, 2011.

\bibitem{Bollinger2001}
M.~V. Bollinger, J.~V. Lauritsen, K.~W. Jacobsen, J.~K. N\o{}rskov, S.~Helveg,
  and F.~Besenbacher.
\newblock One-dimensional metallic edge states in ${\mathrm{mos}}_{2}$.
\newblock {\em Phys. Rev. Lett.}, 87:196803, Oct 2001.

\bibitem{Chen2018}
P.~Chen, W.~W. Pai, Y.-H. Chan, W.-L. Sun, C.-Z. Xu, D.-S. Lin, M.~Y. Chou,
  A.-V. Fedorov, and T.-C. Chiang.
\newblock Large quantum-spin-hall gap in single-layer 1t' wse2.
\newblock {\em Nat. Commun.}, 9(1), May 2018.

\bibitem{Elstner1998}
M.~Elstner, D.~Porezag, G.~Jungnickel, J.~Elsner, M.~Haugk, T.~Frauenheim,
  S.~Suhai, and G.~Seifert.
\newblock Self-consistent-charge density-functional tight-binding method for
  simulations of complex materials properties.
\newblock {\em Phys. Rev. B}, 58(11):7260--7268, 1998.

\bibitem{Foldy1950}
L.~L. Foldy and S.~A. Wouthuysen.
\newblock On the dirac theory of spin \protect{$\frac12$} particles and its
  non-relativistic limit.
\newblock {\em Phys. Rev.}, 78:29--36, 1950.

\bibitem{Gaus2011}
M.~Gaus, Q.~Cui, and M.~Elstner.
\newblock Dftb3: Extension of the self-consistent-charge density-functional
  tight-binding method (scc-dftb).
\newblock {\em J. Chem. Theory Comput.}, 7(4):931--948, 2011.

\bibitem{Gong2014}
Y.~Gong, J.~Lin, X.~Wang, G.~Shi, S.~Lei, Z.~Lin, X.~Zou, G.~Ye, R.~Vajtai,
  B.~I. Yakobson, H.~Terrones, M.~Terrones, B.~K. Tay, J.~Lou, S.~T.
  Pantelides, Z.~Liu, W.~Zhou, and P.~M. Ajayan.
\newblock Vertical and in-plane heterostructures from ws$_2$/mos$_2$
  monolayers.
\newblock {\em Nat. Mater.}, 13(12):1135--1142, 2014.

\bibitem{Grimme2017}
S.~Grimme, C.~Bannwarth, and P.~Shushkov.
\newblock A robust and accurate tight-binding quantum chemical method for
  structures, vibrational frequencies, and noncovalent interactions of large
  molecular systems parametrized for all spd-block elements ($z$ = 1-86).
\newblock {\em J. Chem. Theory Comput.}, 13(5):1989--2009, Apr. 2017.

\bibitem{Hourahine2020}
B.~Hourahine, B.~Aradi, V.~Blum, F.~Bonafé, A.~Buccheri, C.~Camacho,
  C.~Cevallos, M.~Y. Deshaye, T.~Dumitrică, A.~Dominguez, S.~Ehlert,
  M.~Elstner, T.~van~der Heide, J.~Hermann, S.~Irle, J.~J. Kranz,
  C.~K\"{o}hler, T.~Kowalczyk, T.~Kubař, I.~S. Lee, V.~Lutsker, R.~J. Maurer,
  S.~K. Min, I.~Mitchell, C.~Negre, T.~A. Niehaus, A.~M.~N. Niklasson, A.~J.
  Page, A.~Pecchia, G.~Penazzi, M.~P. Persson, J.~Řezáč, C.~G. Sánchez,
  M.~Sternberg, M.~St\"{o}hr, F.~Stuckenberg, A.~Tkatchenko, V.~W.-z. Yu, and
  T.~Frauenheim.
\newblock Dftb+, a software package for efficient approximate density
  functional theory based atomistic simulations.
\newblock {\em J. Chem. Phys.}, 152(12), Mar. 2020.

\bibitem{Huber1979}
K.~P. Huber and G.~Herzberg.
\newblock {\em Constants of diatomic molecules}, pages 8--689.
\newblock Springer US, Boston, MA, 1979.

\bibitem{Jha2022}
G.~Jha and T.~Heine.
\newblock Dftb parameters for the periodic table: Part iii, spin-orbit
  coupling.
\newblock {\em J. Chem. Theory Comput.}, 18(7):4472--4481, June 2022.

\bibitem{Jha2023}
G.~Jha and T.~Heine.
\newblock Spin-orbit coupling corrections for the gfn-xtb method.
\newblock {\em J. Chem. Phys.}, 158(4), Jan. 2023.

\bibitem{Kane2005}
C.~L. Kane and E.~J. Mele.
\newblock Quantum spin hall effect in graphene.
\newblock {\em Phys. Rev. Lett.}, 95:226801, Nov 2005.

\bibitem{Kane2005-2}
C.~L. Kane and E.~J. Mele.
\newblock ${Z}_{2}$ topological order and the quantum spin hall effect.
\newblock {\em Phys. Rev. Lett.}, 95:146802, Sep 2005.

\bibitem{KohnSham1965}
W.~Kohn and L.~J. Sham.
\newblock Self-consistent equations including exchange and correlation effects.
\newblock {\em Phys. Rev.}, 140:A1133--A1138, 1965.

\bibitem{Kohler2007}
C.~Köhler, T.~Frauenheim, B.~Hourahine, G.~Seifert, and M.~Sternberg.
\newblock Treatment of {Collinear} and {Noncollinear} {Electron} {Spin} within
  an {Approximate} {Density} {Functional} {Based} {Method}.
\newblock {\em J. Phys. Chem. A}, 111(26):5622--5629, July 2007.

\bibitem{Liu2021-gb}
Y.~Liu, C.~Zeng, J.~Yu, J.~Zhong, B.~Li, Z.~Zhang, Z.~Liu, Z.~M. Wang, A.~Pan,
  and X.~Duan.
\newblock Moiré superlattices and related moiré excitons in twisted van der
  {Waals} heterostructures.
\newblock {\em Chem. Soc. Rev.}, 50(11):6401--6422, June 2021.
\newblock Publisher: Royal Society of Chemistry (RSC).

\bibitem{Ma20162}
Y.~Ma, L.~Kou, Y.~Dai, and T.~Heine.
\newblock Proposed two-dimensional topological insulator in site.
\newblock {\em Phys. Rev. B}, 94:201104, Nov 2016.

\bibitem{Ma2016}
Y.~Ma, L.~Kou, Y.~Dai, and T.~Heine.
\newblock Two-dimensional topological insulators in group-11 chalcogenide
  compounds:
  ${M}_{2}\mathrm{Te}\phantom{\rule{0.28em}{0ex}}(m=\mathrm{Cu},\phantom{\rule{0.16em}{0ex}}\mathrm{Ag})$.
\newblock {\em Phys. Rev. B}, 93:235451, Jun 2016.

\bibitem{Ma2018}
Y.~Ma, L.~Kou, B.~Huang, Y.~Dai, and T.~Heine.
\newblock Two-dimensional ferroelastic topological insulators in single-layer
  janus transition metal dichalcogenides
  $m\mathrm{SSe}(m=\mathrm{Mo},\mathrm{W})$.
\newblock {\em Phys. Rev. B}, 98:085420, Aug 2018.

\bibitem{Monkhorst1976-tk}
H.~J. Monkhorst and J.~D. Pack.
\newblock Special points for {Brillouin}-zone integrations.
\newblock {\em Phys. Rev.}, 13(12):5188--5192, June 1976.
\newblock Publisher: American Physical Society (APS).

\bibitem{Mulliken1955}
R.~S. Mulliken.
\newblock Electronic population analysis on lcao-mo molecular wave functions.
  i.
\newblock {\em J. Chem. Phys.}, 23(10):1833--1840, 1955.

\bibitem{Naik2018-pl}
M.~H. Naik and M.~Jain.
\newblock Ultraflatbands and shear solitons in moiré patterns of twisted
  bilayer transition metal dichalcogenides.
\newblock {\em Phys. Rev. Lett.}, 121(26):266401, Dec. 2018.
\newblock Publisher: American Physical Society (APS).

\bibitem{Porezag1995}
D.~Porezag, T.~Frauenheim, T.~Köhler, G.~Seifert, and R.~Kaschner.
\newblock Construction of tight-binding-like potentials on the basis of
  density-functional theory: Application to carbon.
\newblock {\em Phys. Rev. B}, 51(19):12947--12957, 1995.

\bibitem{Qi2011}
X.-L. Qi and S.-C. Zhang.
\newblock Topological insulators and superconductors.
\newblock {\em Rev. Mod. Phys.}, 83:1057--1110, Oct 2011.

\bibitem{Qian14}
X.~Qian, J.~Liu, L.~Fu, and J.~Li.
\newblock Quantum spin hall effect in two-dimensional transition metal
  dichalcogenides.
\newblock {\em Science}, 346:1344--1347, 2014.

\bibitem{Seifert1996}
G.~Seifert, D.~Porezag, and T.~Frauenheim.
\newblock Calculations of molecules, clusters, and solids with a simplified
  lcao-dft-lda scheme.
\newblock {\em Int. J. Quantum Chem.}, 58:185--192, 1996.

\bibitem{Tran2019-ct}
K.~Tran, G.~Moody, F.~Wu, X.~Lu, J.~Choi, K.~Kim, A.~Rai, D.~A. Sanchez,
  J.~Quan, A.~Singh, J.~Embley, A.~Zepeda, M.~Campbell, T.~Autry, T.~Taniguchi,
  K.~Watanabe, N.~Lu, S.~K. Banerjee, K.~L. Silverman, S.~Kim, E.~Tutuc,
  L.~Yang, A.~H. MacDonald, and X.~Li.
\newblock Evidence for moiré excitons in van der {Waals} heterostructures.
\newblock {\em Nature}, 567(7746):71--75, Mar. 2019.
\newblock Publisher: Springer Science and Business Media LLC.

\bibitem{Wahiduzzaman2013-sv}
M.~Wahiduzzaman, A.~F. Oliveira, P.~Philipsen, L.~Zhechkov, E.~van Lenthe,
  H.~A. Witek, and T.~Heine.
\newblock {DFTB} parameters for the periodic table: {Part} 1, electronic
  structure.
\newblock {\em J. Chem. Theory Comput.}, 9(9):4006--4017, Sept. 2013.
\newblock Publisher: American Chemical Society (ACS).

\bibitem{Yang2007}
Y.~Yang, H.~Yu, D.~York, Q.~Cui, and M.~Elstner.
\newblock Extension of the self-consistent-charge density-functional
  tight-binding method: Third-order expansion of the density functional theory
  total energy and introduction of a modified effective coulomb interaction.
\newblock {\em J. Phys. Chem. A}, 111(42):10861--10873, 2007.

\end{thebibliography}

\end{document}